\documentclass[twocolumn,trackchanges]{aastex62}
\usepackage{natbib}
\usepackage{amsmath}
\usepackage{amssymb}
\hypersetup{linkcolor=red,citecolor=orange,filecolor=cyan,urlcolor=magenta}

\newcommand{\ang}{\mbox{\AA} }

\newcommand{\galex}{\textit{GALEX}}
\newcommand{\hst}{\textit{HST}}
\newcommand{\jwst}{\textit{JWST}}
\newcommand{\spitzer}{\textit{Spitzer}}
\newcommand{\herschel}{\textit{Herschel}}
\newcommand{\magphys}{\texttt{MAGPHYS}}
\newcommand{\magphysz}{\texttt{MAGPHYS+photo}-$z$}
\newcommand{\magphysagn}{\texttt{MAGPHYS+AGN}}
\newcommand{\lephare}{\texttt{LePhare}}
\newcommand{\zspec}{$z_\mathrm{spec}$}
\newcommand{\zphot}{$z_\mathrm{phot}$}
\newcommand{\zbias}{$z\text{-}\mathrm{bias}$}

\newcommand{\tauv}{\hbox{$\hat{\tau}_{V}$}}

\shorttitle{\magphysz}
\shortauthors{Battisti et al.}

\begin{document}

\title{\magphysz: Constraining the Physical Properties of Galaxies with Unknown Redshifts}

\correspondingauthor{A. J. Battisti}
\email{andrew.battisti@anu.edu.au}

\author[0000-0003-4569-2285]{A. J. Battisti}
\affil{Research School of Astronomy and Astrophysics, Australian National University, Cotter Road, Weston Creek, ACT 2611, Australia}
\affil{ARC Centre of Excellence for All Sky Astrophysics in 3 Dimensions (ASTRO 3D), Australia}

\author[0000-0001-9759-4797]{E. da Cunha}
\affil{Research School of Astronomy and Astrophysics, Australian National University, Cotter Road, Weston Creek, ACT 2611, Australia}
\affil{International Centre for Radio Astronomy Research, University of Western Australia, 35 Stirling Hwy, Crawley, WA 6009, Australia}
\affil{ARC Centre of Excellence for All Sky Astrophysics in 3 Dimensions (ASTRO 3D), Australia}

\author[0000-0002-3247-5321]{K. Grasha}
\affil{Research School of Astronomy and Astrophysics, Australian National University, Cotter Road, Weston Creek, ACT 2611, Australia}
\affil{ARC Centre of Excellence for All Sky Astrophysics in 3 Dimensions (ASTRO 3D), Australia}

\author[0000-0001-7116-9303]{M. Salvato}
\affil{Max-Planck-Institut f\"{u}r extraterrestrische Physik, Giessenbachstrasse 1, D-85748, Garching, Bayern, Germany}

\author[0000-0002-3331-9590]{E. Daddi}
\affil{CEA, IRFU, DAp, AIM, Universit\'e Paris-Saclay, Universit\'e Paris Diderot, Sorbonne Paris Cit\'e, CNRS, F-91191 Gif-sur-Yvette, France}

\author[0000-0003-3085-0922]{L. Davies}
\affil{ICRAR, The University of Western Australia, 35 Stirling Highway, Crawley, WA 6009, Australia}

\author[0000-0002-8412-7951]{S. Jin}
\affil{CEA, IRFU, DAp, AIM, Universit\'e Paris-Saclay, Universit\'e Paris Diderot, Sorbonne Paris Cit\'e, CNRS, F-91191 Gif-sur-Yvette, France}
\affil{School of Astronomy and Space Science, Nanjing University, Nanjing 210093, People’s Republic of China}
\affil{Key Laboratory of Modern Astronomy and Astrophysics, Nanjing University, Nanjing 210093, People’s Republic of China}

\author[0000-0001-9773-7479]{D. Liu}
\affil{Max-Planck-Institut f\"{u}r Astronomie, K\"{o}nigstuhl 17, D-69117 Heidelberg, Germany}

\author[0000-0002-3933-7677]{E. Schinnerer}
\affil{Max-Planck-Institut f\"{u}r Astronomie, K\"{o}nigstuhl 17, D-69117 Heidelberg, Germany}

\author[0000-0002-6748-0577]{M. Vaccari}
\affil{Department of Physics and Astronomy, University of the Western Cape, Robert Sobukwe Road, 7535 Bellville, Cape Town, South Africa}
\affil{INAF - Istituto di Radioastronomia, via Gobetti 101, 40129 Bologna, Italy}
\collaboration{(COSMOS collaboration)}



\begin{abstract}
We present an enhanced version of the multiwavelength spectral modeling code \magphys\ that allows the estimation of galaxy photometric redshift and physical properties (e.g., stellar mass, star formation rate, dust attenuation) simultaneously, together with robust characterization of their uncertainties. The self-consistent modeling over ultraviolet to radio wavelengths in \magphysz\ is unique compared to standard photometric redshift codes. The broader wavelength consideration is particularly useful for breaking certain degeneracies in color vs. redshift for dusty galaxies with limited observer-frame ultraviolet and optical data (or upper limits). We demonstrate the success of the code in estimating redshifts and physical properties for over 4,000 infrared-detected galaxies at $0.4<z<6.0$ in the COSMOS field with robust spectroscopic redshifts. We achieve high photo-$z$ precision ($\sigma_{\Delta z/(1+z_\mathrm{spec})}\lesssim0.04$), high accuracy (i.e., minimal offset biases; $\mathrm{median}(\Delta z/(1+z_\mathrm{spec}))\lesssim0.02$), and low catastrophic failure rates ($\eta\simeq4\%$) over all redshifts. Interestingly, we find that a weak 2175\ang absorption feature in the attenuation curve models is required to remove a subtle systematic \zphot\ offset ($z_\mathrm{phot}-z_\mathrm{spec}\simeq-0.03$) that occurs when this feature is not included. As expected, the accuracy of derived physical properties in \magphysz\ decreases strongly as redshift uncertainty increases. The all-in-one treatment of uncertainties afforded with this code is beneficial for accurately interpreting physical properties of galaxies in large photometric datasets. Finally, we emphasize that \magphysz\ is not intended to replace existing photo-$z$ codes, but rather offer flexibility to robustly interpret physical properties when spectroscopic redshifts are unavailable. The \magphysz\ code is publicly available online.
\end{abstract} 

\section{Introduction}
Obtaining accurate distances to galaxies, typically inferred from cosmological redshift \citep[$z$;][]{hogg99, condon&matthews18}, is an essential first step in any observational study of cosmology or galaxy evolution. This is because the distance is required to derive meaningful physical properties for a galaxy from the observed spectral energy distribution \citep[SED; see][for reviews of SED-fitting]{walcher11, conroy13}. However, due to the observational expense required to obtain accurate redshifts via spectroscopy, \zspec, it is much more practical to estimate redshifts using photometric observations, \zphot. This has spawned numerous `photo-$z$' codes designed to accurately estimate redshifts from photometric data \citep[see recent review by][]{Salvato18}. A majority of these codes employ libraries of spectral models of stellar populations or galaxies (either empirically- or theoretically-based) that are combined in a linear, positive-only combination to determine redshift probabilities based on $\chi^2$ minimization and/or through a Bayesian frame-work \citep[e.g.,][]{benitez00, bolzonella00, leborgne02, ilbert06, rowan-robinson08, brammer08}, although machine learning methods are also being utilized \citep[see][and references therein]{Salvato18}. 

Despite the considerable success of many previously released photo-$z$ codes, a notable concern among them is that the SED templates utilized can often be combined in an arbitrary manner that may not be physically realistic in terms of the properties that would be inferred for the galaxy (e.g., stellar mass ($M_*$), star formation rate (SFR), dust attenuation\footnote{Attenuation is the combination of extinction, scattering of light into the line of sight by dust, and geometrical effects due to the star-dust geometry \citep[e.g.,][]{charlot&fall00, calzetti01}. Extinction is the absorption and scattering of light out of the line of sight by dust and has no dependence on geometry \citep[e.g.,][]{fitzpatrick99, draine03}.}). Indeed, this is the primary reason that most photo-$z$ codes attempt to constrain only the redshift and perhaps a couple other physical properties, if any. A common approach in studies of galaxy evolution is to input the \zphot\ value from photo-$z$ codes into separate codes designed specifically to constrain physical properties \citep[e.g.,][]{daCunha08, leja17, boquien19} and in doing so often ignore entirely the uncertainty of \zphot, or include it through an indirect manner \citep[e.g.,][]{kriek09, johnston15}. Any \zphot\ uncertainty will translate into additional uncertainty in all luminosity-dependent quantities (e.g., $M_*$, SFR, total dust luminosity ($L_\mathrm{dust}$)). Therefore, in this type of approach the uncertainty values for galaxy properties will always be underestimated. 

Another detail worth mentioning is that most photo-$z$ codes utilize data only in the ultraviolet (UV) to near-infrared (NIR\footnote{We adopt the boundary between near-IR and mid-IR to be 5~\micron. This corresponds to the typical wavelength where the dust emission begins to dominate over stellar emission in the SED of a star-forming galaxy.}) wavelength range to constrain the photometric redshift. Extending the wavelength coverage available in the analysis adds additional leverage in determining \zphot\ and can also help resolve various degeneracies that arise in SED fitting \citep[e.g., Lyman vs. Balmer break or $V$-band dust attenuation vs. redshift degeneracy, $A_V$-$z$;][]{dunlop07, daCunha15}. These degeneracies arise because $z$, stellar age, $A_V$, and metallicity can have similar effects on the observed color of the UV-NIR SED. These effects can be disentangled using observational data of the thermal emission from dust in the mid-IR through (sub-)mm wavelength region (or also radio). In addition, assumptions of dust attenuation have a large impact on physical properties derived from SED modeling and therefore it is crucial to incorporate the full spectral range simultaneously for self-consistency whenever possible. As a caveat, it is important to note that active galactic nuclei (AGN) can significantly alter the SEDs of galaxies and this will impact the reliability of derived quantities if the AGN component is not accounted for in the fitting procedure.

We have developed an extension of the widely utilized spectral fitting code Multiwavelength Analysis of Galaxy Physical Properties \citep[\magphys;][]{daCunha08, daCunha15} that enables robust characterization of \zphot\ and numerous physical properties simultaneously in a self-consistent and physically motivated manner. This extension will be referred to throughout the text as ``\magphysz"\footnote{All versions of the \magphys\ code are publicly available online at \url{www.iap.fr/magphys}.}. The advantage of this approach over previous photo-$z$ methods is the consistent incorporation of the \zphot\ uncertainties into the uncertainties of all physical properties that are estimated. Similar to previous versions of the \magphys\ code, \magphysz\ simultaneously models the emission by stellar populations with the attenuation and emission by dust in galaxies based on the assumption of energy balance. We note that similar methodologies have been employed to develop other spectral fitting codes that model the full SED from UV to millimeter (mm) wavelengths \citep[e.g.,][]{leja17, boquien19}, with some also being able to constrain \zphot\ \citep[e.g.,][]{han&han14, chevallard&charlot16}. The reliability of \magphys\ in recovering accurate physical properties is demonstrated in its release paper \citep{daCunha08} and has recently been shown to work well on spatially resolved scales of $\gtrsim$1~kpc \citep{smith&hayward18}. A preliminary version of \magphysz\ was demonstrated in \citet{daCunha15}, however this was limited to the application of a small sample of sub-mm galaxies and is updated considerably in this work to be reliable for a broader range of galaxies. In this paper we outline a description of the code and demonstrate its ability to accurately constrain photometric redshifts and galaxy properties, together with all of their uncertainties, in a robust and reliable manner. 

As a final point, it is worth emphasizing that \magphysz\ is not intended to replace existing stand-alone photo-$z$ codes when the primary parameter of interest is the photometric redshift because those codes run significantly faster and can be `tuned' for higher redshift precision and accuracy than our code. Instead, \magphysz\ is primarily designed for determining physical properties of galaxies with the added flexibility to incorporate photo-$z$ estimates and its uncertainties into the property uncertainties. Additionally, our code is better suited for studying dusty star-forming galaxies for which IR data can provide improved photometric constraints over traditional codes, especially in instances when only upper limits are available at observer-frame UV-optical wavelengths.

This paper is organized as follows: Section~\ref{magphys_method} describes the \magphysz\ code, Section~\ref{data} presents the observed sample used to test the code, Section~\ref{results} outlines the results of applying the code to galaxies in the COSMOS field, Section~\ref{discussion} discusses additional tests of \magphysz\ and a comparison to a commonly adopted photo-$z$ code, and Section~\ref{conclusion} summarizes the main conclusions of the paper. A detailed description of the updated dust attenuation prescription used in \magphysz\ and the characterization of the effective dust attenuation curves for galaxies at $0.1\lesssim z\lesssim3$ will be presented in a companion paper (Battisti et al. in prep). In a Throughout this work we adopt a $\Lambda$-CDM cosmological model, $H_0=70$~km/s/Mpc, $\Omega_M=0.3$, $\Omega_{\Lambda}=0.7$.

\section{\texttt{MAGPHYS+photo-}\texorpdfstring{\MakeLowercase{\textit{z}}}{z_math} Description}\label{magphys_method} 
\magphys\ \citep{daCunha08, daCunha15} is a SED-fitting code designed to self-consistently determine galaxy properties based on an energy-balance approach using rest-frame UV through radio photometry in a Bayesian formalism. We refer the reader to release papers above for a detailed explanation of the \magphys\ code and its methodology. The \magphysz\ code is based on the ``\magphys\ high-$z$" extension presented in \citet{daCunha15}. We outline here a summary of the primary differences between high-$z$ and photo-$z$ versions relative to the original code \citep{daCunha08}:
\begin{itemize}
 \item The prior on the star-formation history (SFH) is modified to be more appropriate for high-$z$ galaxies. This SFH rises linearly at early ages and then declines exponentially, whereas before it was only an exponential declining function (both versions include random bursts of star formation superimposed onto the continuous SFH).
 \item The range of the dust optical depth prior is broader to reflect the higher optical depths observed in high-$z$ galaxies. 
 \item The prior for the equilibrium dust temperature is broader to reflect the wider range in dust temperatures observed in high-$z$ sub-mm galaxies. 
 \item A prescription for intergalactic medium (IGM) absorption in the ultraviolet is added. In the high-$z$ version, the prescription from \citet{madau95} is adopted. For the photo-$z$ version, we adopt the more recent prescription of \citet{inoue14}. 
 \item For the photo-$z$ version, we introduce an additional component in the attenuation curve for the diffuse ISM to characterize additional attenuation due to a 2175\ang feature (details presented below). 
 \end{itemize} 
Regarding the IGM prescription, we tested utilizing both IGM prescriptions on the COSMOS galaxies used in this study and found that it had a minor effect on the values of \zphot for the majority of cases. We attribute the minor differences to the fact that for this dataset the typical uncertainties in the rest-frame far-UV are comparable to or larger than the differences that arise between these two IGM prescriptions \citep[$\sim$10\%; see Figure 4 of][]{inoue14}. In addition, we impose a minimum photometric uncertainty of 10\% to each band (to mitigate over-fitting issues and potential filter zeropoint offsets) that also acts to reduce \zphot\ differences between these prescriptions. However, we note that differences will become more important with increased photometric accuracy and/or finer photometric sampling of the far-UV region. It is also worth noting that the actual IGM absorption will be stochastic in nature, which is not accounted for in our current prescription, and this can introduce small offsets in \zphot\ estimates \citep[e.g.,][]{brinchmann17}. Below we present an overview of the models and methodology implemented in \magphysz. 

The choice to adopt a minimum photometric uncertainty of 10\% to each band warrants additional discussion. The primary motivation in this choice is driven by the fact that systematic offsets are often found between different photometric datasets and are typically of order 10\% (see also Appendix~\ref{app_CANDELS}). Such differences can arise from differences in methodology to obtain the photometry or in adopted flux zeropoints. In some photo-$z$ codes, these effects can be accounted for by adopting corrections for each band that minimize residuals between the photometry and the models for galaxies with spectroscopic redshifts. However, this approach cannot be adopted by \magphysz\ by virtue of our desire to characterize the physical properties because these offsets have the same effect as altering the model SEDs (hence affecting the derived properties). We tested utilizing lower minimum uncertainties and find that this leads to significantly underestimated \zphot\ uncertainties (discussed further in Section~\ref{magphys_photoz}), which in turn leads to corresponding underestimates of property uncertainties relative to the `true' value from standard \magphys\ runs (i.e., fixed to the \zspec\ values; see Section~\ref{compare_prop}).

The stellar emission (dominant in the UV-NIR regime) of each model is calculated such that the luminosity per unit wavelength emerging at time $t$ from a model galaxy is expressed as:
\begin{equation}
L_\lambda^\mathrm{\,em}(t)=\int_0^t dt^\prime \Psi(t-t^\prime)\,l^\mathrm{SSP}_\lambda(t^\prime,Z) \,\exp[-\hat\tau_\lambda(t^\prime)] \,,
\label{llambda}
\end{equation}
where $l^\mathrm{SSP}_\lambda(t^\prime,Z)$ is the luminosity emitted per unit wavelength per unit mass by a simple stellar population (SSP) of age $t^\prime$ and metallicity $Z$, $\Psi(t-t^\prime)$ is the star formation rate evolution with time (i.e. the star formation history), and $\hat\tau_\lambda(t^\prime)$ is the effective absorption optical depth seen by stars of age $t^\prime$. The SSP emission is computed using the spectral population synthesis models of \citet{bruzual&charlot03}, with a \citet{chabrier03} initial mass function. We adopt a uniform prior in metallicity from 0.2 to 2 times solar. 

Nebular emission lines are currently not implemented in \magphys. This should have a minimal impact for general application to most massive galaxies when utilizing broadband photometry (see Section~\ref{broad_inter_photoz}). However, we expect the impact of emission lines to become more important on photomerty in the case of low-mass, high-sSFR galaxies. Implementing nebular emission into \magphys\ and \magphysz\ is planned for a forthcoming release.

The dust attenuation, as defined by the `effective' absorption optical depth \citep{charlot&fall00} of the dust as seen by stars, $\hat{\tau}_\lambda$, is given by:
\begin{equation}\label{eq:tau_definition}
\hat{\tau}_\lambda(t^\prime)=
\begin{cases}
\hat{\tau}_\lambda^{\,\mathrm{BC}}+\hat{\tau}_\lambda^{\,\mathrm{ISM}} \, & \text{for $t^\prime\leq t_\mathrm{BC}$,} \\
\hat{\tau}_\lambda^{\,\mathrm{ISM}}\, & \text{for $t^\prime> t_\mathrm{BC}$}\,,
\end{cases}
\end{equation}
where $\hat{\tau}_\lambda^{\,\mathrm{BC}}$ is the effective attenuation optical depth of dust in stellar birth clouds, $\hat{\tau}_\lambda^{\,\mathrm{ISM}}$ is the effective attenuation optical depth in the diffuse ISM, and $t_\mathrm{BC}\simeq10^7$~yr. 

For the diffuse ISM attenuation curve, we utilize an additional component to account for attenuation from a 2175\ang feature. The 2175\ang absorption bump is a prominent feature in the Milky Way (MW) extinction curve \citep[e.g.,][]{draine03}, but this feature is typically much weaker in strength or completely absent in dust attenuation curves\footnote{The additional scattering and geometric effects at play in dust attenuation act to reduce the overall strength of the 2175\ang feature relative to extinction curves \citep[e.g.,][]{gordon97, gordon00, witt&gordon00, seon&draine16}} \citep[e.g.,][]{calzetti94, noll09a, wild11, buat12, kriek&conroy13, reddy15, scoville15, salmon16, battisti17b, salim18} and for this reason it is often excluded in SED modeling \citep{charlot&fall00, daCunha08}. We find that this feature is necessary to reduce systematic residuals between the observations and models in standard \magphys, with an average bump strength of $\sim$30\% the MW value being necessary for  IR-detected galaxies from $0.1\lesssim z\lesssim3$ (Battisti et al. in prep.). As will be discussed in Section~\ref{2175_zbias}, we also find that the lack of a parameter accounting for additional attenuation from the 2175\ang feature introduces a subtle systematic underestimate of \zphot\ in \magphysz. Some photo-$z$ codes allow for possible 2175\ang absorption by considering fits based on multiple attenuation/extinction curves, which generally acts to improve the quality of fits \citep[e.g.,][]{ilbert09}. It is also worth noting that the need for this feature is sometimes avoided in other photo-$z$ codes through the use of zeropoint offsets or template corrections, which are not performed in \magphysz.

Here we briefly summarize implementation of the 2175\ang absorption feature that will be fully detailed in Battisti et al. (in prep.). We make the assumption that this feature follows a behavior similar to that of the MW extinction curve, which is well characterized using a Lorentzian-like Drude profile \citep[e.g.,][]{fitzpatrick&massa07}:
\begin{equation}\label{eq:drude}
D(E_b,\lambda) = \frac{E_b(\lambda\,\Delta\lambda)^2}{(\lambda^2-\lambda_0^2)^2+(\lambda\,\Delta\lambda)^2} \,,
\end{equation}
where $\lambda_0$ is the central wavelength of the feature, $\Delta\lambda$ is its FWHM, and $E_b$ is an amplitude constant that defines the bump strength. The average MW extinction curve has values of $\lambda_0=2175.8$~\AA, $\Delta\lambda=470$~\AA, and $E_b=3.3$ \citep{fitzpatrick99}. This $E_b$ value refers to the amplitude of the Drude profile in terms of the total-to-selective extinction curve $k_\lambda\equiv A_\lambda/E(B-V)$. However for our purposes, the Drude profile is defined in terms of normalized optical depth, $\tau_\lambda/\tau_V$ \citep{charlot&fall00}, which we denote as $E_b'$ to avoid confusion. The relationship between the two versions of the bump strength is the following: 
\begin{equation}\label{eq:ebprime}
E_b' = E_b/R_V^{\,\mathrm{ISM}} \,,
\end{equation}
where $R_V^{\,\mathrm{ISM}}$ is the total-to-selective attenuation in the $V$ band from the diffuse ISM (note that $R_V=k_V$). For comparison to dust curves that do not utilize two components (birth cloud and ISM), one can simply adopt $R_V^{\,\mathrm{ISM}}=R_V$. For example, the MW extinction curve has a value of $E_b'=1.06$ \citep[$R_V=3.1$;][]{fitzpatrick99}. 

\begin{figure}
\begin{center}
$\begin{array}{cc}
\includegraphics[width=0.225\textwidth]{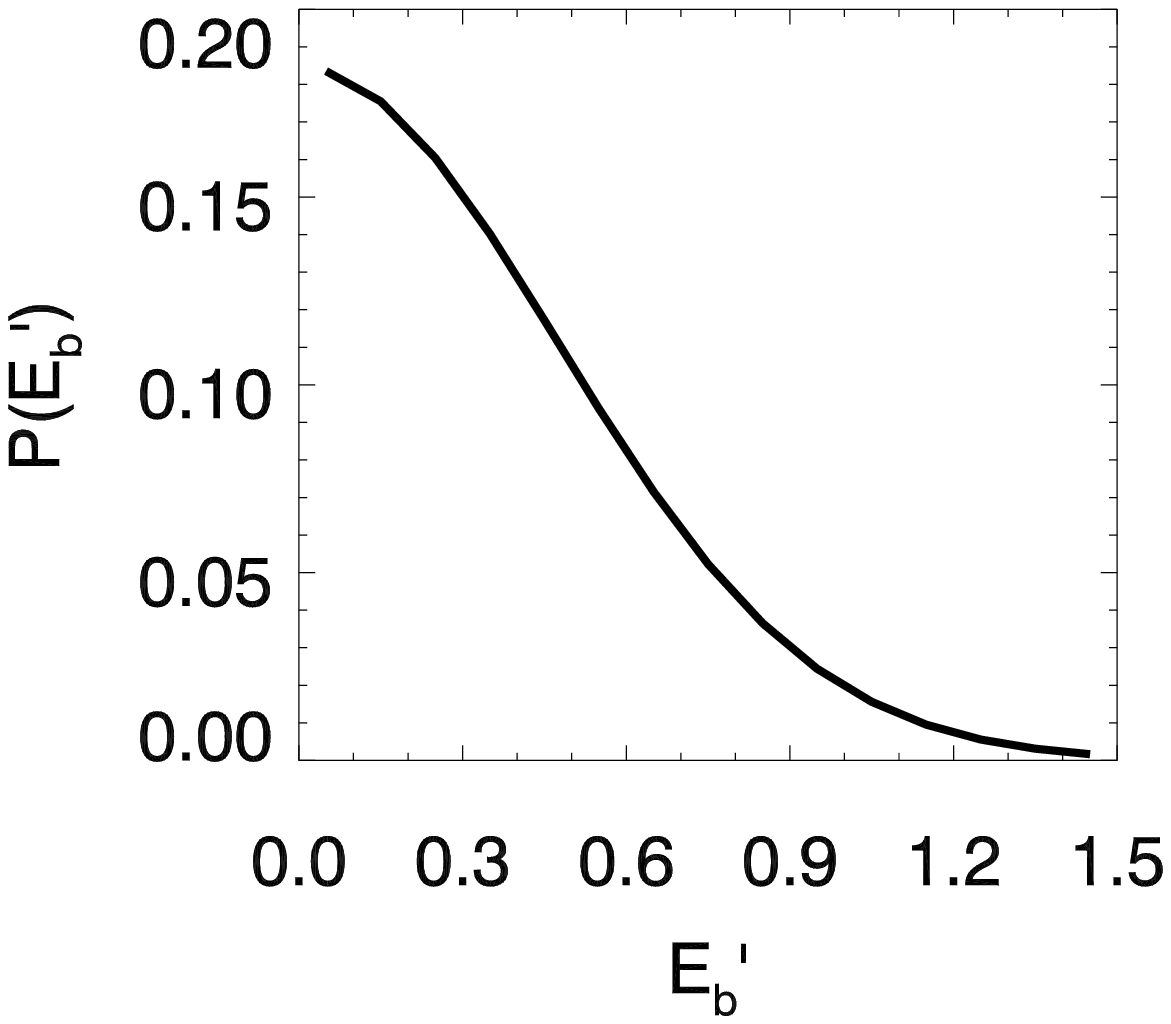} &
\includegraphics[width=0.225\textwidth]{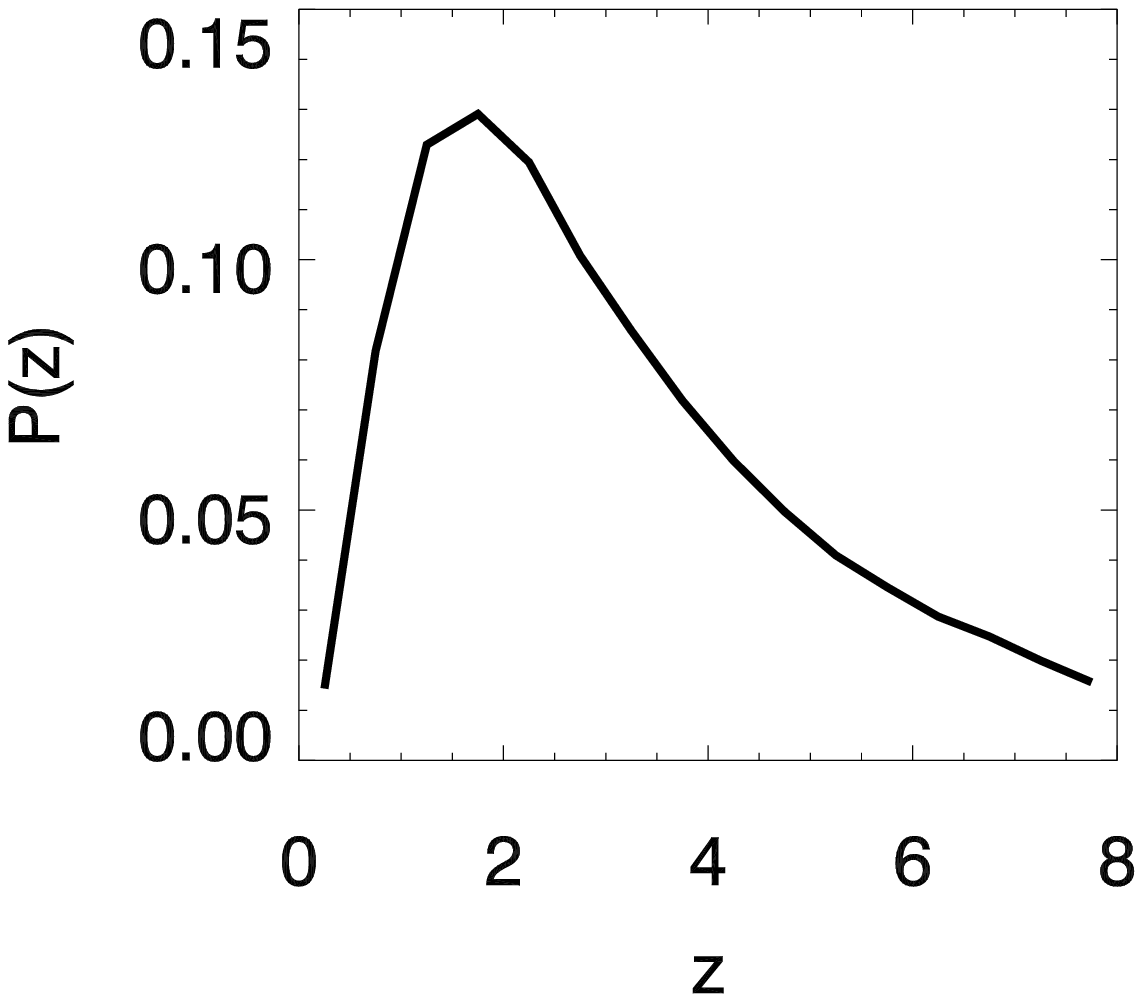} \\
\end{array}$
\end{center}
\vspace{-0.5cm}
\caption{Prior  for the 2175\ang feature bump strength ($E_b'$, see eq~\ref{eq:drude} and \ref{eq:ebprime}; Battisti et al. in prep.) and redshift \citep[$z$;][]{daCunha15} utilized in \magphysz. The prior distributions for all other parameters are the same as outlined in \citet{daCunha15}.
\label{fig:bump_z_prior}}
\end{figure}

The attenuation curve of the birth clouds remains unchanged from previous versions of \magphys, 
\begin{equation}
\hat{\tau}_\lambda^{\,\mathrm{BC}}=(1-\mu)\,\hat\tau_V\,(\lambda/5500\ang)^{-1.3} \,,
\end{equation}
where \tauv\ is the effective $V$-band optical depth seen by stars younger than $t_\mathrm{BC}$ in the birth clouds and $\mu$ is the fraction of \tauv\ seen by stars older than $t_\mathrm{BC}$ (i.e., stars in the diffuse ISM component, $\mu=\hat\tau_V^{\,\mathrm{ISM}}/(\hat\tau_V^{\,\mathrm{BC}}+\hat\tau_V^{\,\mathrm{ISM}})$). The updated attenuation curve characterizing the diffuse ISM is:
\begin{equation}\label{eq:tau_ISM}
\hat{\tau}_\lambda^{\,\mathrm{ISM}}=\mu\,\hat\tau_V\,[(\lambda/5500\ang)^{-0.7}+D(E_b',\lambda)] \,,
\end{equation}
where, the central wavelength and FWHM of the 2175\ang feature are fixed to the MW value and only the amplitude defining the bump strength, $E_b'$, is free to vary within the code. The distribution of the prior for $E_b'$ was chosen based on the observed strength inferred from intermediate-band data of galaxies in COSMOS (Battisti et al. in prep.) and is shown in Figure~\ref{fig:bump_z_prior}.

The dust emission models (dominant in the mid-IR-sub-mm regime) combine contributions from four components: PAHs and hot dust emitting in the mid-IR, and warm and cold dust components in thermal equilibrium that emit in the far-IR to sub-mm. For the energy balance process within the code, these four components are subdivided into those associated with the birth cloud component (PAHs+hot dust+warm dust) and those associated with the diffuse ISM component (PAHs+hot dust+warm dust+cold dust) and are required to match the energy lost by the attenuation in each component at UV-NIR wavelengths. 

The radio component is modeled based on assuming a far-infrared/radio correlation \citep[$q$;][]{Yun01}, and fixed slopes and relative contributions for the thermal and non-thermal radio emission (following \citealt{dale&helou02}; see \citealt{daCunha15} for details). For the analysis in this paper, we do not include radio data because the assumption of a fixed $q$ with redshift will result in the radio data having little power in constraining \zphot\ when far-IR data are also available, as is the case for our chosen samples. However, it is possible to include radio data and it will generally act to improve the overall constraints of various physical properties in \magphysz , especially when far-IR data are unavailable.

The predicted fluxes of the user-specified filter set for each model (stellar and dust+radio) are then calculated for redshifts in the range $0.1<z<8.0$. The assumed prior for the redshift distribution of the models is shown in Figure~\ref{fig:bump_z_prior}. This redshift prior is optimized to avoid oversampling unlikely regions of redshift space for application to deep far-IR/sub-mm surveys for which the distribution of galaxy number counts as a function of redshift has a similar shape \citep[e.g.,][]{zavala14, daCunha15}. We note that the chosen redshift prior is not well suited for application of \magphysz\ to very low-redshift ($z\lesssim0.4$) or very high-redshift ($z\gtrsim7.5$) galaxies because very few models are generated at these redshifts. As a result, any runs that return \zphot\ values in these redshift regimes will be unreliable and we recommend that they not be utilized in subsequent analysis. Our model stellar ages are required to be younger than the age of the Universe at a given redshift. For each redshift, the predicted model flux at each band is computed by first applying the appropriate IGM absorption prescription to the stellar emission at that redshift and then convolving the total model SED (stellar+dust+radio; matched via energy balance) in the observed-frame with the filter response functions. We then compare the observed fluxes of our galaxies in all the observed bands with the predicted model fluxes by computing the $\chi^2$ goodness-of-fit for each model in our library \citep[defined in][]{daCunha08}. Upper limits are included in the $\chi^2$ estimate by setting the flux to zero and adopting the upper limit value as the flux uncertainty. 

Finally, we build likelihood distributions of each parameter in our model, including the redshift, by marginalizing the probability of each model $P \propto \exp(-\chi^2/2)$ over all other model parameters. We take our estimates of each parameter to be the median of its likelihood distribution, and the 1$\sigma$ confidence range to be the 16th to 84th percentile range.

\section{Data and Measurements}\label{data}
To characterize the reliability of \magphysz , we must utilize a large sample of galaxies with spectroscopically derived redshifts (\zspec) over a large redshift range that we can compare to our derived values of \zphot. For the purpose of this code it is also desirable to use a galaxy sample with extensive multi-wavelength photometric data that spans the full range from UV to sub-mm. For our analysis, we utilize the COSMOS field \citep{scoville07, capak07} because it meets all of these requirements and is unparalleled in terms of the breadth of spectroscopic and multi-wavelength photometric data that are available from numerous telescope facilities. This field has a wealth of IR photometry available, which allows full exploitation of the multiwavelength modeling capability of \magphys. We use two photometric+spectroscopic catalogs of galaxies in the COSMOS field that are optimized for different redshift ranges: the first are part of the Galaxy And Mass Assembly (GAMA) survey and the second are produced by the COSMOS team. Below we provide a brief description of these catalogs and our method for selecting the galaxy samples used for the analysis.

\subsection{GAMA - G10 Sample}
The GAMA survey compiled a highly complete multiwavelength photometric \citep{driver11} and spectroscopic sample of galaxies \citep{baldry10, robotham10, hopkins13} covering 280 deg$^2$ especiall in three equatorial (G09, G12 and G15) and two southern (G02 and G23) regions, as well as a subset of the COSMOS field (G10), although the latter utilizes previously obtained spectroscopy from the zCOSMOS survey \citep{lilly07, lilly09}. For this study we use only the G10 region, for which the photometric catalog is described in \citet{andrews17} and the spectroscopic catalog is described in \citet{davies15}. 

The catalog of \citet{andrews17} contains data from the \galex, Canada-France-Hawaii telescope (CFHT), Subaru, UltraVista, \spitzer, and \herschel\ that span the full UV to sub-mm wavelength range and is summarized in Table~\ref{tab:filt_extinct}. The photometry is aperture-matched utilizing the Lambda-Adaptive Multi-Band Deblending Algorithm in R \citep[\texttt{LAMBDAR};][]{wright16} algorithm. The provided photometry is already corrected for Galactic extinction for all bands from FUV to $K_s$ using $E(B-V)$ values from the \citet{schlegel98} Milky Way dust maps and the extinction curve values listed in \citet[][see their Table 3]{andrews17}.

For our parent sample of spectroscopic galaxies, we follow the recommended selection criteria of $\mathtt{Z\_BEST}>0.0001$,  $\mathtt{Z\_USE}<3$, and  $\mathtt{STAR\_GALAXY\_CLASS}=0$ \citep{davies15}. This selection provides a reduced sample of 20,364 objects. We further refine this sample to only consider galaxies at $z_\mathrm{spec}>0.4$ with $S/N>3$ (signal-to-noise) in any \spitzer/MIPS, or \herschel\ band (i.e., requiring at least one detection in the mid-IR to sub-mm wavelength range), which leaves 2,542 galaxies at $0.4<z\lesssim1.6$. We note that the IR-detection criteria tends to select dustier galaxies, on average, in the spectroscopic sample, but that the amount of reddening ($A_V$) can vary significantly (also, $A_V$ is typically larger for higher redshifts, and vice versa). A test of the code on galaxies without this restriction is presented in Section~\ref{LePhare_comparison}.
 
\subsection{COSMOS2015 \& Super-Deblended Sample}
To extend the redshift range in our analysis, we utilize the latest COSMOS master spectroscopic catalog (curated by M. Salvato for internal use within the COSMOS collaboration), together with the COSMOS2015 \citep{laigle16} and ``Super-deblended'' \citep{jin18} photometric catalogs. These catalogs combine data from numerous surveys (see release papers for individual surveys utilized) in a homogeneous manner. The COSMOS2015 catalog provides photometry for the UV through mid-IR wavelength range from the \galex, CFHT, Subaru, UltraVista, and \spitzer/IRAC. The COSMOS2015 photometric catalog is corrected for Galactic extinction for all bands from NUV to $K_s$ using the provided $E(B-V)$ values based on the \citet{schlegel98} Milky Way dust maps and adopting the same extinction curve values as used in G10 for consistency \citep{andrews17}. We adopt total flux values (3\arcsec + aperture correction) using the prescription described in \citet{laigle16}, because these are better suited for combining with the total flux measurements of the IR data in the Super-deblended catalog \citep{jin18}. The Super-deblended catalog extends the wavelength coverage of COSMOS sources for the mid-IR through radio wavelength range from the \spitzer/MIPS, \herschel, SCUBA2, AzTEC, MAMBO, and Very Large Array (VLA). \citet{jin18} use near-IR and radio priors to overcome severe \textit{Herschel} sub-mm source blending issues and achieve deeper detections \citep[framework described in][]{liu18}. The bands utilized from the COSMOS2015+Super-deblended catalogs are summarized in Table~\ref{tab:filt_extinct}. These catalogs have also been cross-matched with X-ray sources from the Chandra COSMOS catalogs \citep{elvis09, civano12, civano16, marchesi16} and the XMM/Newton Wide-Field Survey in the COSMOS Field \citep{hasinger07, brusa07, brusa10, cappelluti09}. For brevity, we will refer to the combined COSMOS2015+Super-deblended catalogs as ``C15+SD catalog" throughout the remainder of the paper. 

For our parent sample of spectroscopic galaxies, we utilize cases with robust spectroscopic redshifts  \citep[Quality flag $\mathtt{Q_f}=3$ or 4, as defined in zCOSMOS;][]{lilly09}. The COSMOS master catalog contains many duplicate observations and these are remedied in the manner described below. First, if all duplicates have \zspec\ within $0.01$ of each other, we simply adopt the first value and remove all other duplicate cases, as these differences are below the expected precision of our photo-$z$ solutions. If the \zspec\ of the duplicates disagree by more than $0.01$, we give preference to cases of `very secure redshifts' ($\mathtt{Q_f}=4$) over `secure redshifts' ($\mathtt{Q_f=3}$). In instances of disagreement between cases with similar $\mathtt{Q_f}$, we resolve this in the following manner: 1) duplicate cases with \zspec\ solutions within 0.05 of each other are given a higher ranking relative to single outliers (here we consider all \zspec\ regardless of $\mathtt{Q_f}$ to determine ranking of higher $\mathtt{Q_f}$ cases); and 2) higher spectral resolution ($R$) observations are ranked higher than lower resolution observations (with grism data having the lowest ranking). If after considering these two options there are still duplicates of equal rank, we rank newer observations higher than older observations. After resolving duplicate cases, we are left with a parent sample of 34,785 galaxies. We further refine this sample to only consider galaxies at $z_\mathrm{spec}>1.0$ with $S/N>3$ in any band from \spitzer/MIPS, \herschel, SCUBA, AZTEC, or MAMBO, which leaves 2,133 galaxies at $1\leq z<6$. For situations where the same galaxy is present in both the G10 and C15+SD catalogs, we adopt the G10 photometry because the \texttt{LAMBDAR} approach is designed to provide self-consistent photometry over the full UV to sub-mm range \citep{wright16}. This removes 130 IR-detected galaxies from the C15+SD catalog, leaving a sample of 2,003 galaxies.


\begin{table}
\caption{COSMOS filter set and Galactic extinction curve values.}
\begin{tabular}{lcccc}
\hline
Band  & Facility/Instr. & $\lambda_\mathrm{eff}(\micron)$ & $k(\lambda)^a$ & Catalog$^b$\\
\hline
FUV  & $GALEX$ & 0.1516 & 8.376 & G10 \\
NUV & $GALEX$ & 0.2267 & 8.741 & G10, C15 \\
$u^{*}$ & CFHT/Mega-Prime & 0.3750 & 4.690 & G10, C15 \\
IA427 & Subaru/SC& 0.42635 & 4.260 & G10, C15 \\
$B$ & Subaru/SC& 0.4460 & 4.039 & G10, C15 \\
$g'$ & Subaru/SC& 0.4480 & 3.738 & G10 \\
IA464 & Subaru/SC& 0.46351 & 3.843 & G10, C15 \\
IA484 & Subaru/SC& 0.48492 & 3.621 & G10, C15 \\
IA505 & Subaru/SC& 0.50625 & 3.425 & G10, C15 \\
IA527 & Subaru/SC& 0.52611 & 3.264 & G10, C15 \\
$V$ & Subaru/SC& 0.5484 & 3.147 & G10, C15 \\
IA574 & Subaru/SC& 0.57648 & 2.937 & G10, C15 \\
IA624 & Subaru/SC& 0.62329 & 2.694 & G10, C15 \\
$r'$ & Subaru/SC& 0.6295 & 2.586 & G10, C15 \\
IA679 & Subaru/SC& 0.67811 & 2.430 & G10, C15 \\
IA709 & Subaru/SC& 0.70736 & 2.289 & G10, C15 \\
IA738 & Subaru/SC& 0.73615 & 2.150 & G10, C15 \\
$i'$ & Subaru/SC& 0.7641 & 1.923 & G10, C15 \\
IA767 & Subaru/SC& 0.76849 & 1.996 & G10, C15 \\
IA827 & Subaru/SC& 0.82445 & 1.747 & G10, C15 \\
$z'$ & Subaru/SC& 0.9037 & 1.436 & G10 \\
$z^{++}$ & Subaru/SC& 0.9037 & 1.436 & C15 \\
$Y$ & Subaru/HSC & 0.9791 & 1.298 & C15 \\
$Y$ & VISTA/VIRCAM & 1.021 & 1.211 & G10, C15 \\
$J$ & VISTA/VIRCAM & 1.254 & 0.871 & G10, C15 \\
$H$ & VISTA/VIRCAM & 1.646 & 0.563 & G10, C15 \\
$K_s$ & VISTA/VIRCAM & 2.149 & 0.364 & G10, C15 \\
ch1 & IRAC/$Spitzer$ & 3.550 & \nodata & G10, C15 \\
ch2 & IRAC/$Spitzer$ & 4.493 & \nodata & G10, C15 \\
ch3 & IRAC/$Spitzer$ & 5.731 & \nodata & G10, C15 \\
ch4 & IRAC/$Spitzer$ & 7.872 & \nodata & G10, C15 \\
24\micron & MIPS/$Spitzer$ & 23.68 & \nodata & G10, SD \\
70\micron & MIPS/$Spitzer$ & 71.42 & \nodata & G10 \\
100\micron & PACS/$Herschel$ & 100 & \nodata & G10, SD \\
160\micron & PACS/$Herschel$ & 160 & \nodata & G10, SD \\
250\micron & SPIRE/$Herschel$ & 250 & \nodata & G10, SD \\
350\micron & SPIRE/$Herschel$ & 350 & \nodata & G10, SD \\
500\micron & SPIRE/$Herschel$ & 500 & \nodata & G10, SD \\
850\micron & SCUBA-2/JCMT & 850 & \nodata & SD \\
1.1mm & AzTEC/ASTE & 1100 & \nodata & SD \\
1.2mm & MAMBO-2/IRAM & 1200 & \nodata & SD \\
\hline
\end{tabular} \\
$^a$ Extinction curve values adopted as in Table 3 of \citet[][and references outlined therein]{andrews17}\\
$^b$ Denotes the catalog(s) that include this band. G10=GAMA/G10 catalog \citep{andrews17}, C15=COSMOS2015 catalog \citep{laigle16}, SD=Super-deblended catalog \citep{jin18}
\label{tab:filt_extinct}
\end{table}

\subsection{AGN Identification}\label{AGN_ID}
We identify and remove AGN from the main analysis because current public versions of \magphys\ are intended only for purely star-forming galaxies (i.e., it does not include AGN contribution to the SEDs) and therefore properties derived for AGN may be incorrect. Introducing AGN models into \magphys\ is planned for a forthcoming release \citep[\magphysagn; application of a preliminary version is shown in][]{chang17}. AGN are identified using several techniques, including: 1) the \spitzer/IRAC color selections of \citet{donley12}; 2) the \spitzer-\herschel\ color selections of \citet{kirkpatrick13}; 3) the radio-NIR color selection of \citet{seymour08}; and 4) sources with any X-ray detection. We note that for the color selection criteria each photometric band used in a given color is required to have signal-to-noise ratio of $S/N>3$. These methods identify 179 AGN (7.0\%) in the G10 sample and 117 AGN (5.8\%) in the C15+SD sample. This leaves a final sample of 2,363 and 1,886 galaxies in the G10 and C15+SD catalogs, respectively, for our analysis. 


In practice it is not always feasible to identify AGN when limited data is available. In addition, even when various selection methods can be implemented, they can often miss some of the AGN population. As a result, we separately discuss the reliability of the \zphot\ estimates for our AGN in Appendix~\ref{app_AGN}.

\section{Application of \texttt{MAGPHYS+photo-}\texorpdfstring{\MakeLowercase{\textit{z}}}{z_math} to COSMOS}\label{results}
\subsection{Photometric Redshifts}\label{magphys_photoz}

\begin{figure*}[ht]
\begin{center}
\includegraphics[width=1.\textwidth]{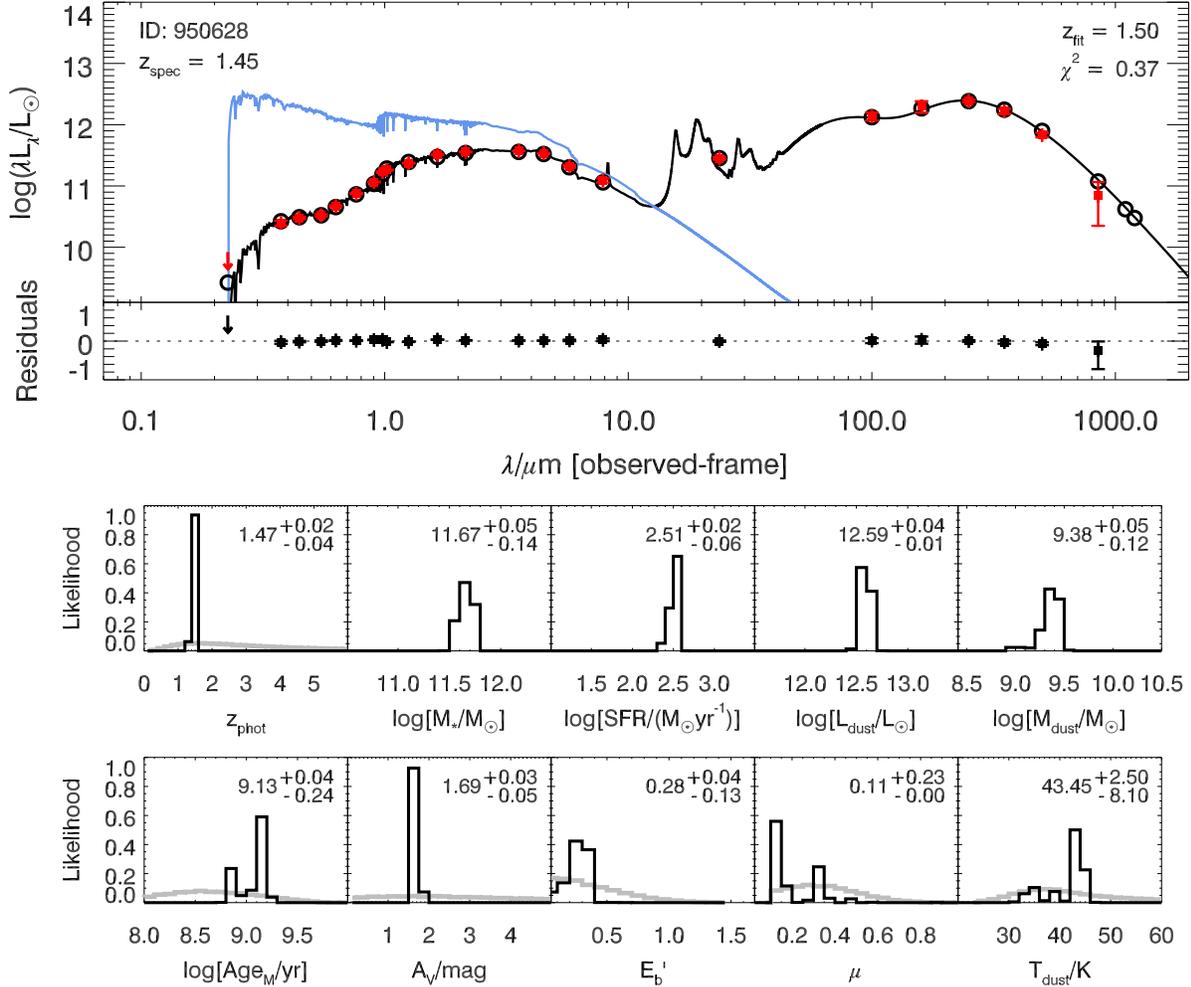}
\end{center}
\caption{\textit{Top:} The upper panel demonstrates the best-fit model from \magphysz\ (black line) on broadband data of a randomly selected galaxy in the C15+SD catalog with excellent far-IR sampling. The filled red squares and open black circles are the observed and model photometry, respectively. The uncertainty on the data are shown by the red error bars. The blue line is the predicted unattenuated stellar population SED for the best-fit model. The lower panel shows the residual between the observed and model photometry. \textit{Bottom:} The two rows show the full likelihood distributions of \zphot , stellar mass, star formation rate, dust luminosity, dust mass, mass-weighted stellar age, $V$-band dust attenuation, 2175\ang bump strength ($E_b'$), the fraction of \tauv\ seen by stars older than $t_\mathrm{BC}$ ($\mu$), and the effective dust temperature, respectively. For parameters with priors, we show the assumed prior distribution as a gray histogram. The top right corner of each panel indicates the median-likelihood value and the 16th-84th percentile range.
\label{fig:magphys_example_COSMOS}}
\end{figure*}

We perform \magphysz\ runs on all galaxies in our GAMA/G10 and C15+SD samples using two filter sets: 1) using only the broadband filters, and 2) using both the broad- and intermediate-band filters (see Table~\ref{tab:filt_extinct}). The choice of using both of these filter sets allows us to examine the influence of using different filters on the results. An example of a \magphysz\ run on the broadband data for a galaxy in the C15+SD sample is shown in Figure~\ref{fig:magphys_example_COSMOS}, together with likelihood distributions for some of the derived properties.

\subsubsection{Using Only Broadband Data}\label{broad_photoz}
\begin{figure*}
\begin{minipage}{0.6\textwidth}
\includegraphics[width=\textwidth]{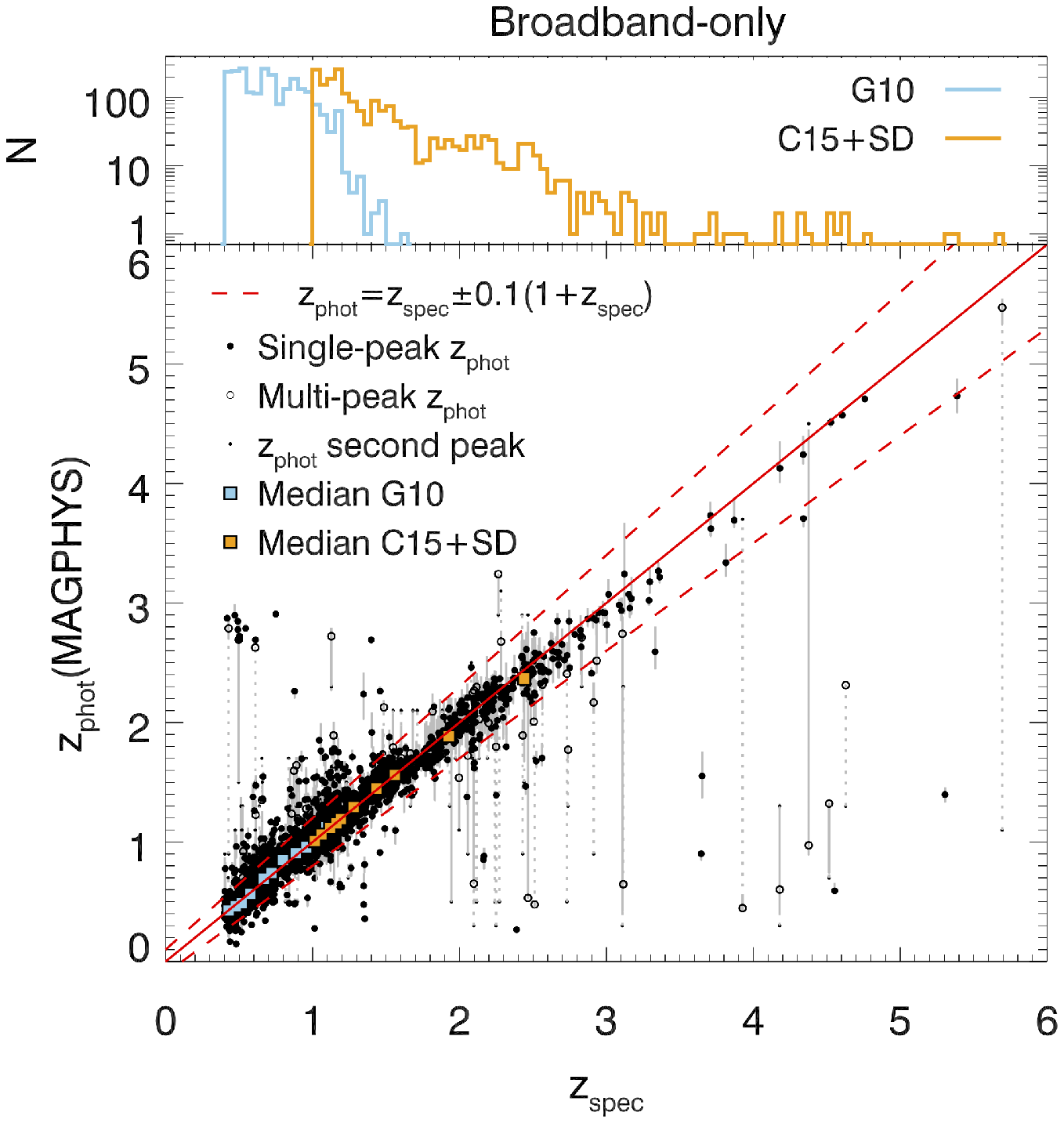}
\end{minipage}
\hspace{3mm}
\begin{minipage}{0.35\textwidth}
\includegraphics[width=\textwidth]{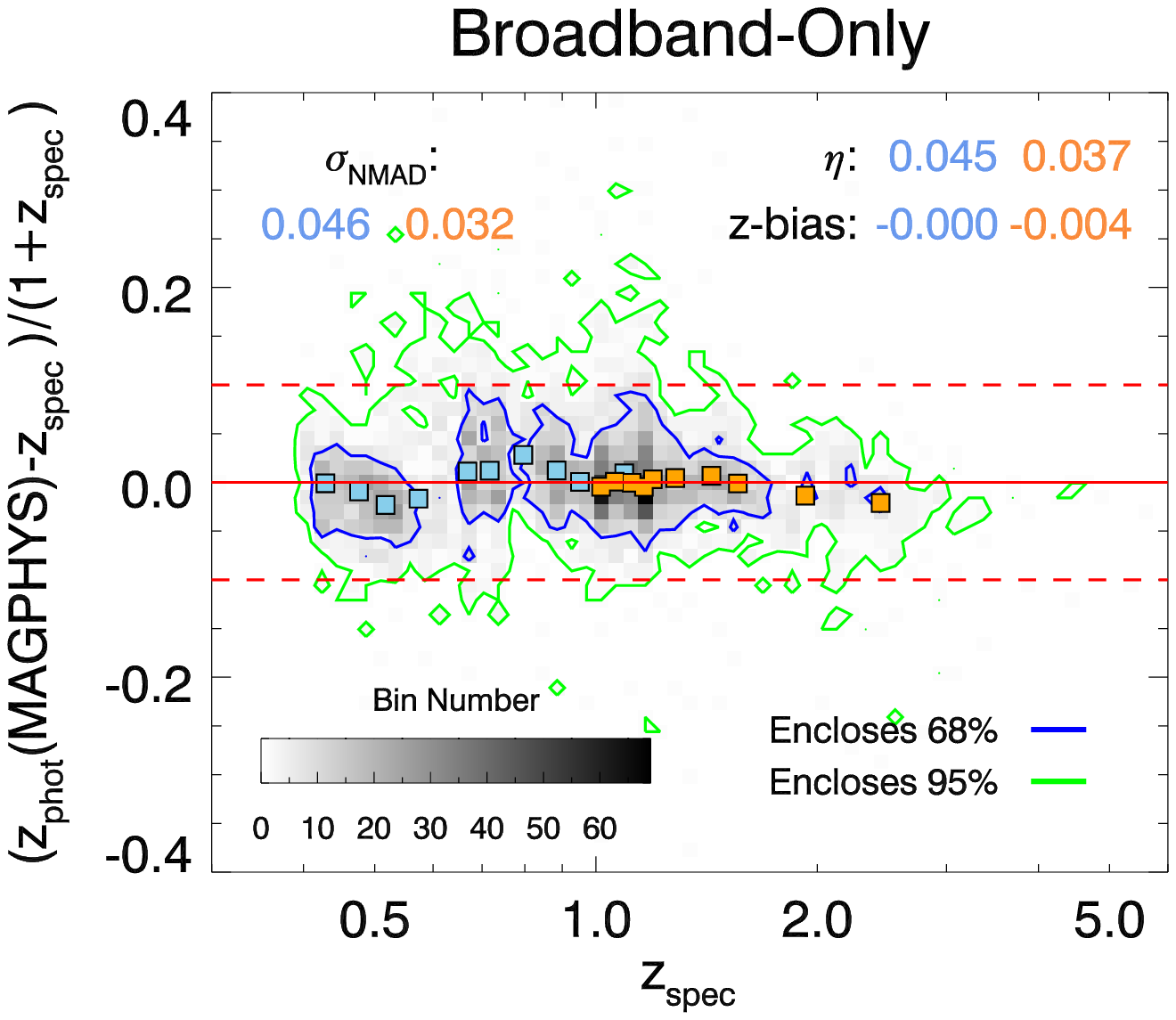}
\\[3mm]
\includegraphics[width=\textwidth]{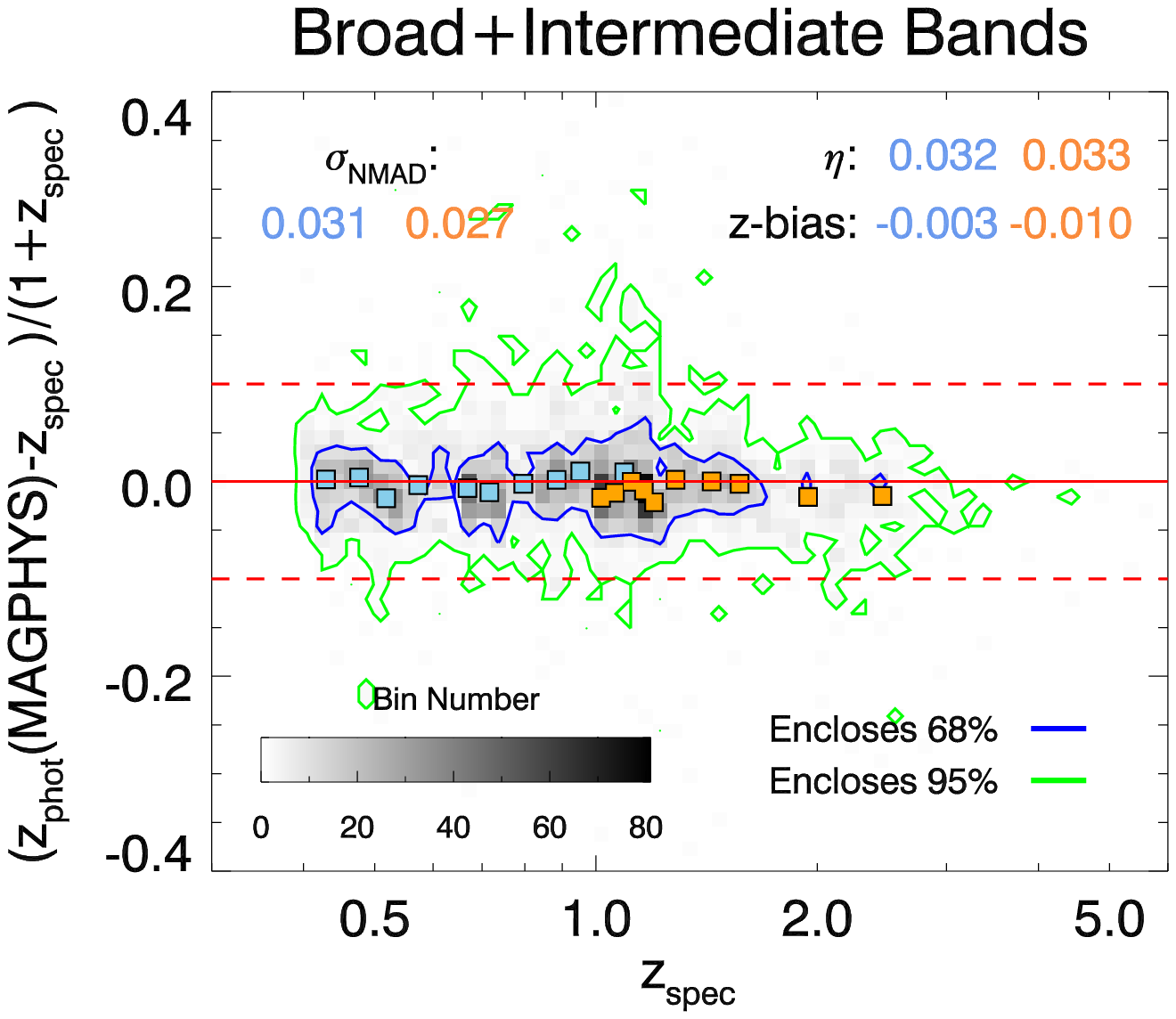}
\end{minipage}
\caption{\textit{Left:} Comparison between the photometric and spectroscopic redshifts for IR-detected galaxies using \magphysz\ on broadband data (circles). The $1\sigma$ confidence range of \zphot\ for each galaxy is represented by a gray error bar. Cases with multiple \zphot\ peaks have the primary redshift peak indicated by open circles and a small filled circles to indicate secondary peaks (when the secondary peak is beyond the $1\sigma$ range they are connected by dotted gray lines). The median values of equal-number bins (excluding catastrophic failures) for the GAMA/G10 and C15+SD samples are shown as light blue and orange squares, respectively. \textit{Top Right:} Redshift accuracy ($\Delta z$/(1+\zspec)) as a function of \zspec\ using broadband data shown in terms of surface density. The redshift scatter ($\sigma_\mathrm{NMAD}$), catastrophic failure rate ($\eta$), and redshift bias (median($\Delta z/(1+z_\mathrm{spec})$)) values are shown for each sample at the top of each panel. There is minimal redshift bias at all redshifts examined among these samples. \textit{Bottom Right:} Redshift accuracy using broad+intermediate band data. There is improvement in $\sigma_\mathrm{NMAD}$ when including intermediate bands, but not in the other metrics. Comparing the top and bottom panels (e.g., at $z_\mathrm{spec}\simeq0.7$) demonstrates that small redshift biases can be dependent on the filter set. \label{fig:photz_vs_specz}}
\end{figure*}

The photometric vs. spectroscopic redshift distribution of the full G10 and C15+SD samples using broadband-only data are shown in Figure~\ref{fig:photz_vs_specz}, \textit{Left}. Cases where the \zphot\ distribution has $>2\%$ of its probability located at a secondary peak value (i.e., degenerate \zphot\ solutions) are flagged as being ``multi-peak \zphot .'' These cases are considered less reliable and we make this distinction for the purpose of subsequent analysis. We find that the fraction of multi-peak \zphot\ cases for our G10 and C15+SD samples are 1.5\% and 3.1\%, respectively. We also denote the median values for both samples and it can be seen that the overall agreement between the \zphot\ and \zspec\ values are excellent. 

Here we define some commonly adopted metrics for characterizing the performance of photo-$z$ results and give the corresponding values found for \magphysz\ using only broadband data. The precision, or scatter of the data, is characterized using the normalized median absolute deviation \citep[NMAD;][]{hoaglin83}, defined as 
\begin{equation}\label{eq:sigma_NMAD}
\sigma_\mathrm{NMAD}=1.48\times \mathrm{median} \left(\left\lvert \frac{\Delta z-\mathrm{median}(\Delta z)}{1+z_\mathrm{spec}} \right\rvert\right) \,,
\end{equation}
where $\Delta z=z_\mathrm{phot}-z_\mathrm{spec}$. The precision of the G10 and C15+SD samples are $\sigma_\mathrm{NMAD}=0.046$ and $0.032$, respectively. We define a catastrophic failure, also referred to as an outlier, as a source with $\Delta z/(1+z_\mathrm{spec})>0.15$. The fraction of catastrophic failures, 
\begin{equation}\label{eq:eta}
\eta=N(\Delta z/(1+z_\mathrm{spec})>0.15)/N(\mathrm{total}) \,,
\end{equation}
for the G10 and C15+SD samples are $\eta=4.5\%$ and $3.7\%$, respectively. The accuracy of the code is quantified as the systematic deviation, or bias, from $z_\mathrm{phot}=z_\mathrm{spec}$ using the median value of the population not considered to be catastrophic failures,
\begin{multline}\label{eq:zbias}
z\text{-}\mathrm{bias} = \mathrm{median}(\Delta z/(1+z_\mathrm{spec}))  \\ \mathrm{for} \,\, \Delta z/(1+z_\mathrm{spec})<0.15 \,. 
\end{multline}
The redshift-normalized median value for the G10 and C15+SD samples are $z\text{-}\mathrm{bias}=0.000$  and $-0.004$, respectively, and are below the achieved precision ($\sigma_\mathrm{NMAD}$). The values of the main metrics for each catalog are summarized in Table~\ref{tab:magphysz_metrics}. We find that $\Delta z/(1+z_\mathrm{spec})<0.05$ for 69\% and 81\% of the G10 and C15+SD samples, respectively. The fraction increase to 91\% and 94\%, respectively, for cases with $\Delta z/(1+z_\mathrm{spec})<0.1$. 

The fraction of cases for which the \zphot\ 1$\sigma$ confidence ranges are in agreement with \zspec\ is 31\% and 62\% for the G10 and C15+SD samples, respectively (58\% and 85\% within 2$\sigma$ agreement). This indicates that we underestimate the \zphot\ uncertainties for our galaxies, particularly for the G10 sample at lower redshifts. Underestimation of \zphot\ uncertainties is a common problem among photo-$z$ codes \citep[e.g.,][]{dahlen13}. In our case, we believe this effect is mainly due to over-fitting the extensive number of available filters with high $S/N$ (despite imposing a minimum 10\% photometric uncertainty to each band) where systematic effects in the photometry can be problematic \citep[e.g., zeropoint issues;][]{dahlen13}. At lower redshifts, there are also fewer models for comparison in our library relative to higher redshifts, due to the redshift prior utilized, and this can impact the uncertainty estimates. However, we note that the fraction of cases with 1$\sigma$ agreement as a function of redshift for \magphysz\ are comparable to other photo-$z$ codes that do not utilize redshift priors (see Section~\ref{LePhare_comparison}). The error underestimation does not significantly impact the estimates for the physical properties (or their uncertainties) because these cases are typically within $\Delta z/(1+z_\mathrm{spec})<0.05$, where \zphot\ uncertainties are less influential on parameter estimates.

\begin{table}
\begin{center}
\caption{Summary of photo-$z$ metrics (defined in Section~\ref{broad_photoz}) for \magphysz\ on COSMOS galaxies.}
\begin{tabular}{lccc}
\hline
 & \multicolumn{3}{c}{Broadband-Only} \\
Catalog & $\sigma_\mathrm{NMAD}$ & $\eta$ & $z\text{-}\mathrm{bias}$ \\
\hline
G10  		  & 0.046 & 0.045 & 0.000 \\
C15+SD  & 0.032 & 0.037 & -0.004 \\
\hline \\
\hline
 & \multicolumn{3}{c}{Broad+Intermediate Bands} \\
Catalog  & $\sigma_\mathrm{NMAD}$ & $\eta$ & $z\text{-}\mathrm{bias}$ \\
\hline
G10         & 0.031 & 0.032 & -0.003 \\
C15+SD  & 0.027 & 0.033 & -0.010 \\
\hline
\end{tabular}
\label{tab:magphysz_metrics}
\end{center}
\end{table}

Among all catastrophic failure cases, we find that 15\% and 21\% of them are the `primary redshift' of multi-peak \zphot\ cases (open circle) for G10 and C15+SD, respectively. For reference, we find that the primary redshift of multi-peak \zphot\ cases are in agreement (below the catastrophic threshold) for 53\% and 75\% of these cases in G10 and C15+SD, respectively. This indicates that the multi-peak \zphot\ cases, which make up 2.1\% of our total population, have a higher rate of failure than the single peaked solutions, as would be expected, although the reliability of the primary redshift is still reasonable. We also examine the best-fit $\chi^2$ values as an indicator of \zphot\ reliability. For this we consider the distribution of $\chi^2$ for each sample, which depends on the filter set and data quality, and fit a Gaussian distribution to the lower 90\% of each population to determine its mean, $\bar{\chi}^2$, and dispersion, $\sigma(\chi^2)$. These fits give $\bar{\chi}^2_\mathrm{G10}=1.35$ and $\sigma(\chi^2_\mathrm{G10})=0.62$, and $\bar{\chi}^2_\mathrm{C15+SD}=0.51$ and $\sigma(\chi^2_\mathrm{C15+SD})=0.23$, where the lower values in the C15+SD sample are primarily due to the lower $S/N$ of the higher redshift data. As expected, we find that the failure rate is higher for cases with high $\chi^2$ values, with 30\% and 34\% of catastrophic failures having $\chi^2>\bar{\chi}^2+4\sigma(\chi^2)$ for G10 and C15+SD, respectively. For comparison, the instance rate of $\chi^2>\bar{\chi}^2+4\sigma(\chi^2)$ for our total population is 4.3\% and 6.3\% for G10 and C15+SD, respectively. High $\chi^2$ values may be indicative of objects for which the current \magphys\ models are not able to reproduce the observations, such as might be expected if fitting single stars (that might contaminate the catalog) or AGN (see Appendix~\ref{app_AGN}). We recommend users of the code to be cautious of \magphysz\ results with multi-peaked \zphot\ distributions and/or particularly high $\chi^2$ values relative to the rest of the population because the results for these cases are likely to be less reliable.

\begin{figure*}
\begin{center}
\includegraphics[width=0.95\textwidth,clip=true]{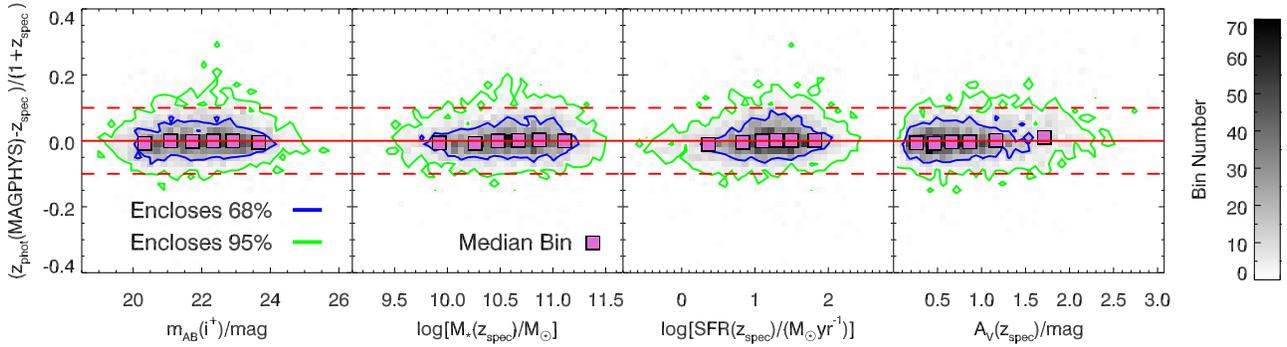}
\end{center}
\vspace{-0.4cm}
\caption{Redshift accuracy ($\Delta z$/(1+\zspec)) for the entire sample (G10 \& C15+SD) as a function of Subaru $i^+$ apparent magnitude, which is related to the $S/N$ quality, and the median values of $M_*$, SFR, and $A_V$ derived using \magphys\ high-$z$ (i.e., fixed to \zspec). The median values of equal-number bins (excluding catastrophic failures) are shown as purple squares. The redshift scatter ($\sigma_\mathrm{NMAD}$), catastrophic failure rate ($\eta$), and redshift bias (median($\Delta z/(1+z_\mathrm{spec})$)) values are similar over the ranges examined and no significant trends are observed.
\label{fig:zphot_vs_param_2Dhist}}
\end{figure*}

We also examine the behavior of the \zbias\ with photometric quality and galaxy physical properties to identify if there are regimes where the code may be less reliable. In Figure~\ref{fig:zphot_vs_param_2Dhist}, we show the \zbias\ for our combined sample as a function of Subaru $i^+$ apparent magnitude, which is related to the $S/N$ quality, and the median values of $M_*$, SFR, and $A_V$ derived using \magphys\ high-$z$ (i.e., fixed to \zspec). We adopt Subaru $i^+$ to allow comparison to \zphot\ results presented in \citet{laigle16} for the COSMOS2015 catalog (see also Section~\ref{LePhare_comparison}). We do not observe any significant trends in \zbias\ with the  $i^+$ apparent magnitude and find comparable dispersions ($\sigma_\mathrm{NMAD}\simeq0.04$) and catastrophic failure rates ($\eta\simeq4\%$) for all bins. For the comparison with physical properties, we adopt fixed-\zspec\ values because $M_*$ and SFR are luminosity dependent (i.e., they depend on \zphot) and the \zspec\ values are a better indication of the `true' galaxy property (see Section~\ref{compare_prop}). Similar to before, we do not observe significant trends in \zbias\ as a function of $M_*$, SFR, and $A_V$. The comparison with $A_V$ indicates that the accuracy does not change substantially with the total amount of dust attenuation, although the precision is dependent on the $S/N$. The results of this Section demonstrate that \magphysz\ is successful in obtaining reliable \zphot\ values over a wide range of redshifts and physical properties.

\subsubsection{Using Broad+Intermediate Band Data}\label{broad_inter_photoz}
Now, we consider \magphysz\ runs that include the broad- and intermediate-band data. As a reminder, emission lines are not included in our models and these can impact observation in narrower filters. For this reason, users should be cautious when utilizing intermediate-band data in \magphysz. However, doing this allows us to examine how much the \zphot\ results change when they are included and also allows for direct comparison to \zphot\ values determined from an independent code that uses these filters (Section~\ref{LePhare_comparison}). The comparison between the photo-$z$ results for the broadband-only and broad+intermediate band fits is shown in Figure~\ref{fig:photz_vs_specz}, \textit{Right}, and includes the metrics that are obtained for each case at the top of each panel. As expected, the precision ($\sigma_\mathrm{NMAD}$) shows an improvement when the intermediate bands are included thanks to the better SED sampling. The catastrophic failure rate ($\eta$) also shows a small decrease for both samples. However, the \zbias\ becomes slightly worse. Another feature worth noting is that the redshift offsets of the samples as function of \zspec\ between these filter sets are slightly different (e.g., at $z_\mathrm{spec}\simeq0.7$ the broadband-only has a slight positive \zbias\ and the broad+intermediate band case is slightly negative). This is not surprising given that the \zphot\ constraints are dependent on which regions of the SED are sampled and how well the filters sample strong features (in this instance the intermediate bands increase the sampling of the Balmer break). There are a couple slight \zbias\ dips (e.g., at $z\sim1$ and $z\sim1.2$) that arise for the broad+intermediate-band fits due to high equivalent width emission lines (in our case arising from [OII](3727\AA)) influencing the intermediate-band photometry, which is currently not accounted for in the models. 

The current lack of emission lines in the models may decrease the level of \zphot\ precision that can be achieved \citep[higher $\sigma_\mathrm{NMAD}$; e.g.,][]{ilbert09} but does not significantly affect the accuracy (\zbias) in most cases, although exceptions to the latter may occur for sources with very high equivalent width lines \citep[high specific-SFR,][]{pacifici15}. In many instances, sources influenced by strong emission lines tend to have multi-peak \zphot\ solutions in \magphysz . For now, we recommend users test the effect of removing bands from a fit if they are suspected to be significantly influenced by an emission line (e.g., if a prominent emission line falls on a particular filter at one of the \zphot\ peaks, try rerunning the code for that source without the filter included (\texttt{fit=0} in the filter file) to see if there are significant changes).

\subsection{Physical Properties and Their Uncertainties}\label{compare_prop}
In this section we outline the ability of \magphysz\ to reliably constrain galaxy physical properties and their uncertainties. As a baseline for this assessment, we compare the properties derived for each galaxy to those of \magphys\ high-$z$ runs where the redshift is fixed to \zspec. We expect these fixed-redshift runs to provide the best estimate for the `true' physical properties. For consistency between the \magphys\ high-$z$ and \magphysz\ versions, we adopt the IGM prescription of \citet{inoue14} instead of \citet{madau95} and also include the 2175\ang feature component in the \magphys\ high-$z$ runs (Battisti et al. in prep.). We utilize the exact same priors for all parameters in both version. All fitting results in these sections use the broadband-only data for the GAMA/G10 and C15+SD samples.

\begin{figure*}
\begin{center}
$\begin{array}{cc}
\includegraphics[width=0.45\textwidth,clip=true]{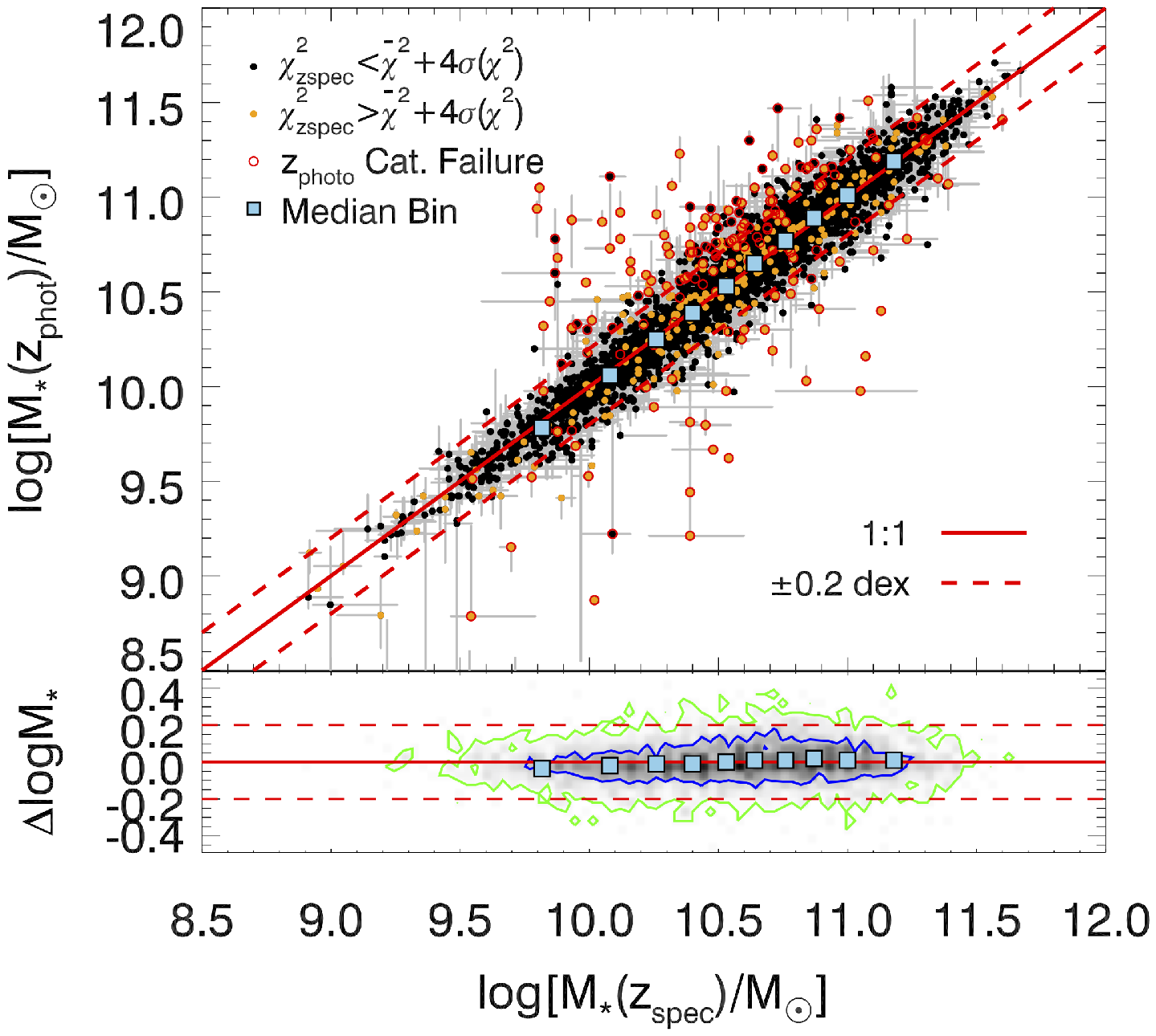} &
\includegraphics[width=0.45\textwidth,clip=true]{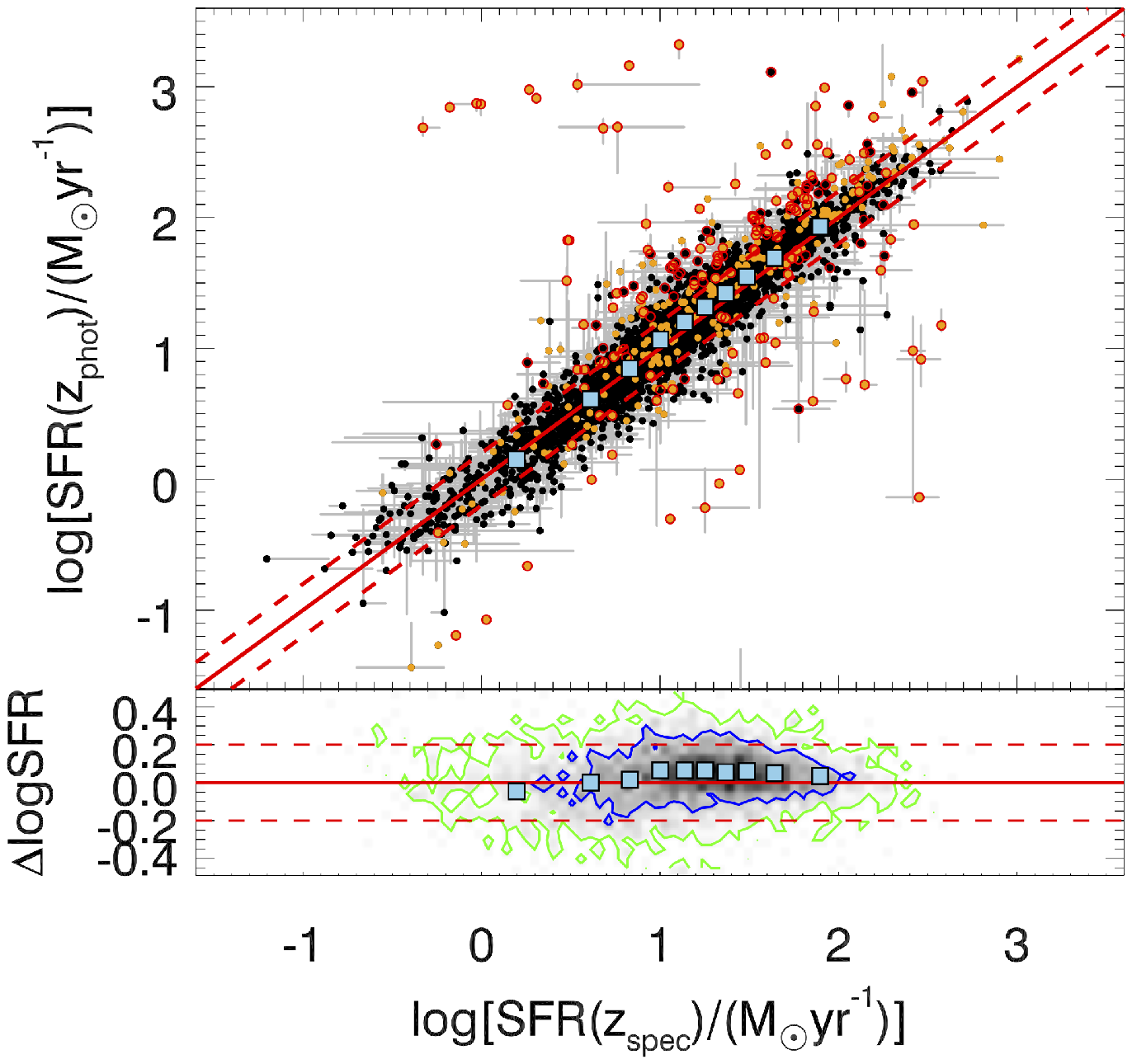} \\
\includegraphics[width=0.45\textwidth,clip=true]{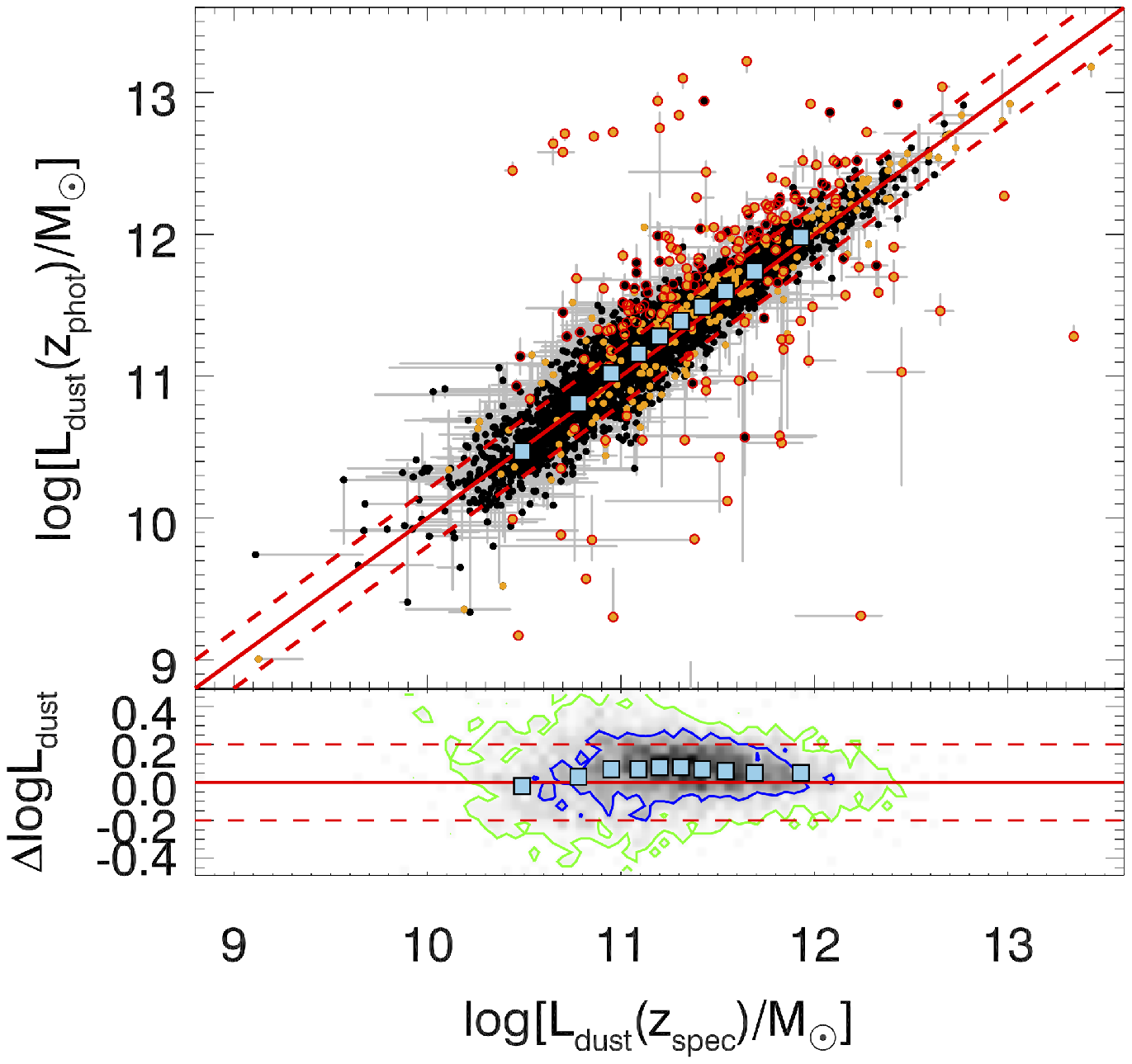} &
\includegraphics[width=0.45\textwidth,clip=true]{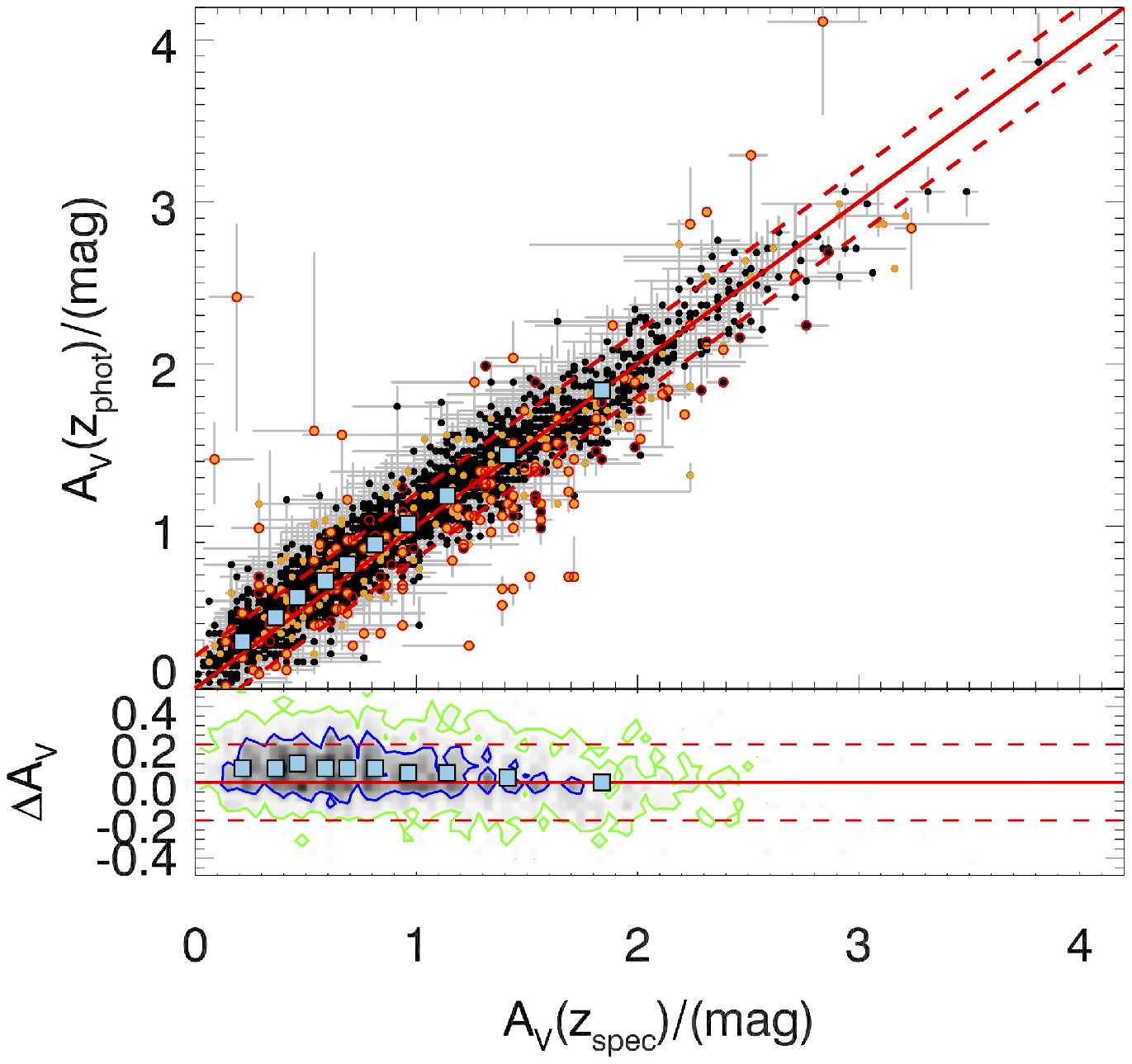} \\
\end{array}$
\end{center}
\vspace{-0.3cm}
\caption{Comparison of stellar mass ($M_*$; \textit{top left}), SFR (\textit{top right}), total dust luminosity ($L_\mathrm{dust}$; \textit{bottom left}), and effective $V$-band dust attenuation ($A_V$; \textit{bottom right}) derived from \magphysz\ and \magphys\ high-$z$ (i.e., redshift fixed to \zspec). Galaxies with relatively poor quality fits ($\chi^2>4$) in \magphys\ high-$z$ (orange circles) or catastrophic failures in \magphysz\ (open red circles) are considered unreliable. For well-fit galaxies (black circles) the agreement is quite good, with median differences that are below the dispersion and with $\sim70\%$ of values agreeing within their 1$\sigma$ confidence range (see Section~\ref{compare_prop}). The surface density of the median difference (photo-$z$ $-$ high-$z$) is shown at the bottom of each panel.
\label{fig:parameter_compare}}
\end{figure*}

The comparison between \magphysz\ and \magphys\ high-$z$ for the derived values of the $M_*$, SFR, $L_\mathrm{dust}$, and $A_V$ are shown in Figure~\ref{fig:parameter_compare}. We distinguish galaxies with poor quality fits in \magphys\ high-$z$, defined as $\chi_{z\mathrm{spec}}^2>\bar{\chi}^2+4\sigma(\chi^2)$, because these cases are likely to have unreliable property estimates for comparison. These high-$\chi^2$ cases comprise 3.7\% and 8.6\% of the G10 and C15+SD samples, respectively. We also distinguish galaxies with \zphot\ catastrophic failures in \magphysz\ because they have drastically different parameter estimates due to the different assumed distances. It is not surprising that many of the cases with poor quality fits in \magphys\ high-$z$ are also catastrophic failures in \magphysz. Below we summarize the results of comparing well-fit galaxies. We find good agreement between parameter values from each code, with a median difference between the photo-$z$ and high-$z$ values of 0.00, 0.04, 0.05, and 0.05 for log($M_*$), log(SFR), log($L_\mathrm{dust}$), and $A_V$, respectively. These values are below the dispersion of the difference, which is 0.10, 0.15, 0.14, and 0.13, respectively. The parameter estimates from both codes are consistent (overlapping within their 1$\sigma$ confidence range) for 82\%, 73\%, 70\%, and 74\%, respectively. 

\begin{figure}
\begin{center}
\includegraphics[width=0.3\textwidth,clip=true]{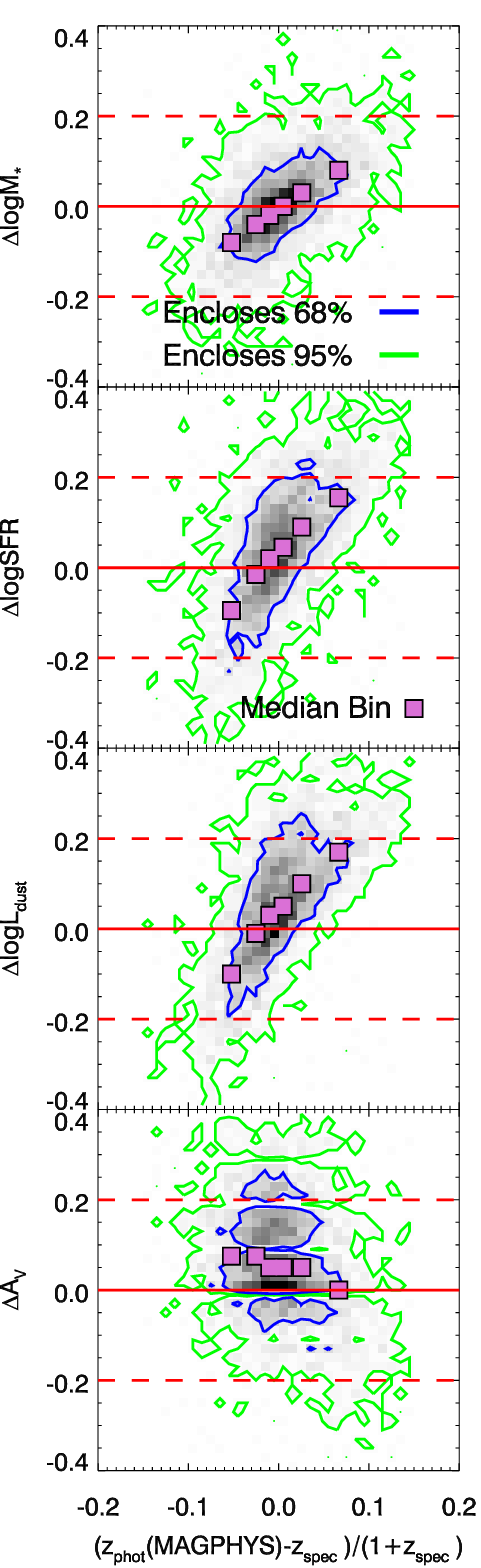}
\end{center}
\vspace{-0.4cm}
\caption{Comparison of the difference in $M_*$, SFR, $L_\mathrm{dust}$, and $A_V$ (photo-$z$ $-$ high-$z$) as a function of the difference between \zphot\ and \zspec. As expected, the largest differences occur for luminosity-dependent quantities.
\label{fig:delta_param_vs_delta_z}}
\end{figure}

The largest differences in parameter estimates between the two versions of the code occur from differences between \zphot\ and \zspec, as would be expected, and this is demonstrated in Figure~\ref{fig:delta_param_vs_delta_z}. The positive trends for $M_*$, SFR, and $L_\mathrm{dust}$ with the redshift offset are a direct result of these quantities being luminosity-dependent (i.e., an overestimated $z$ will infer larger quantities, and vice versa). The slight negative trend with $A_V$ is a result of the degeneracy of this quantity with \zphot, because both of these quantities can have similar effects on the observed colors. 

\begin{figure*}
\begin{center}
$\begin{array}{cc}
\includegraphics[width=0.45\textwidth]{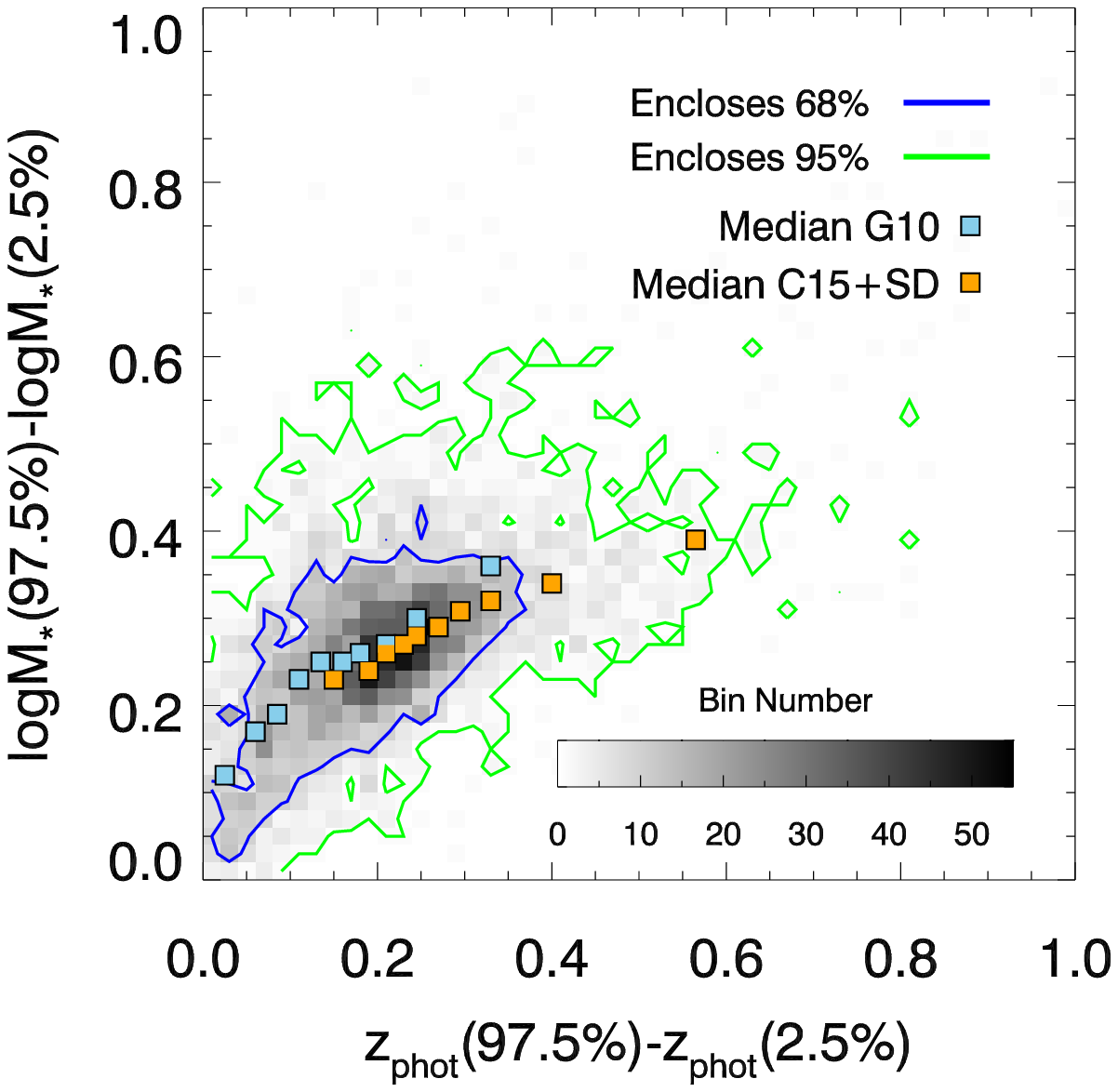} &
\includegraphics[width=0.45\textwidth]{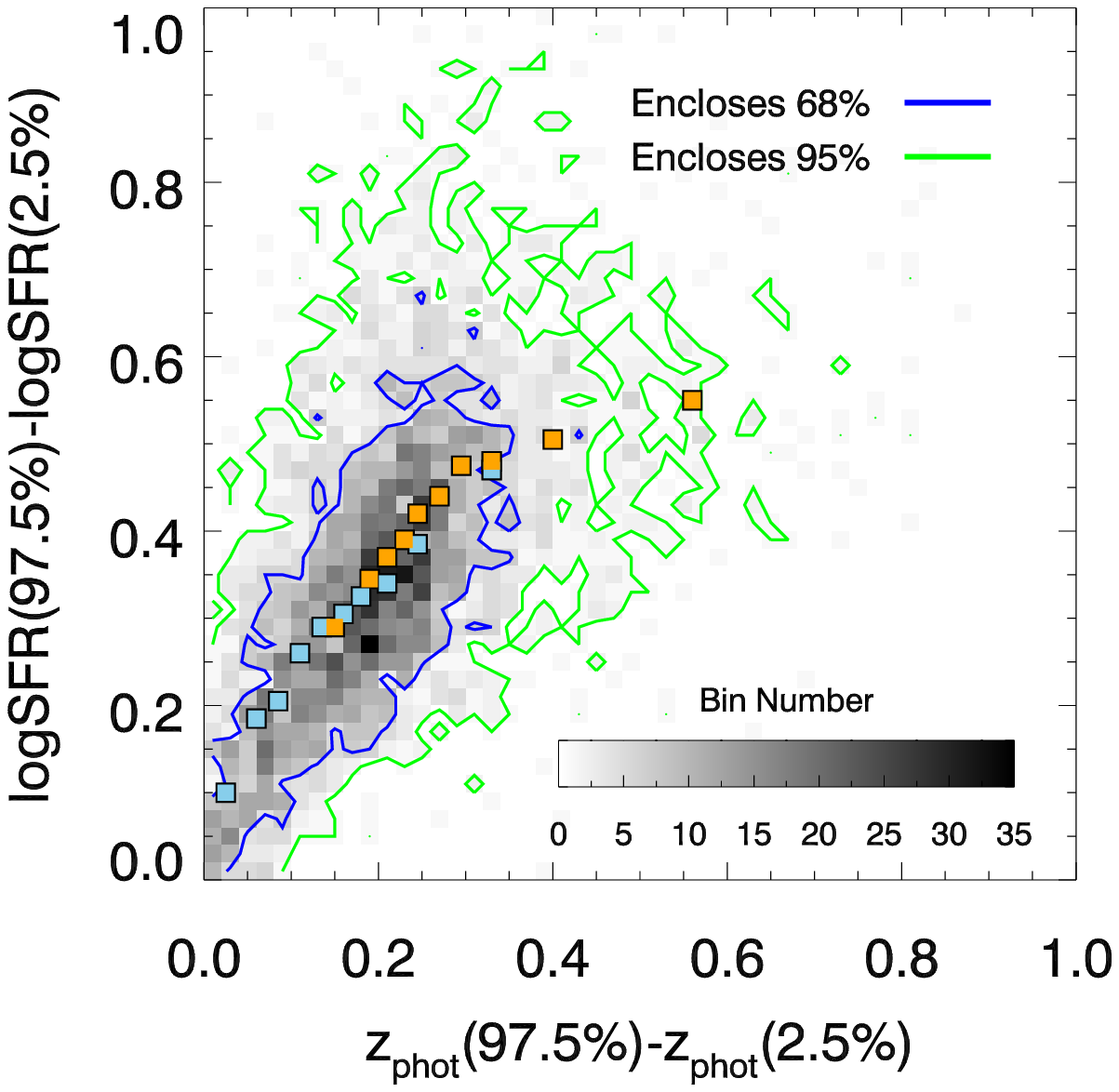} \\
\includegraphics[width=0.45\textwidth]{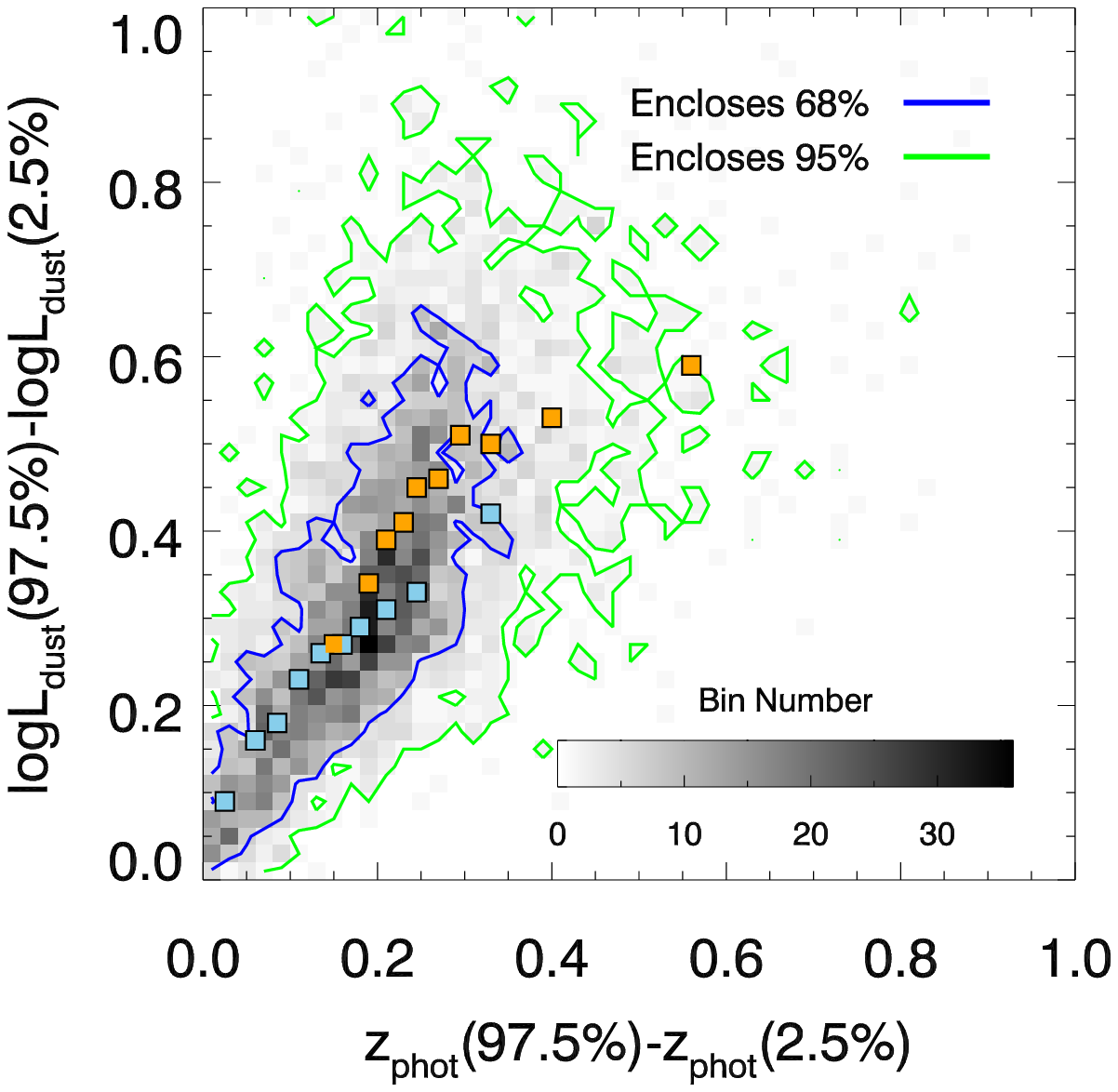} &
\includegraphics[width=0.45\textwidth]{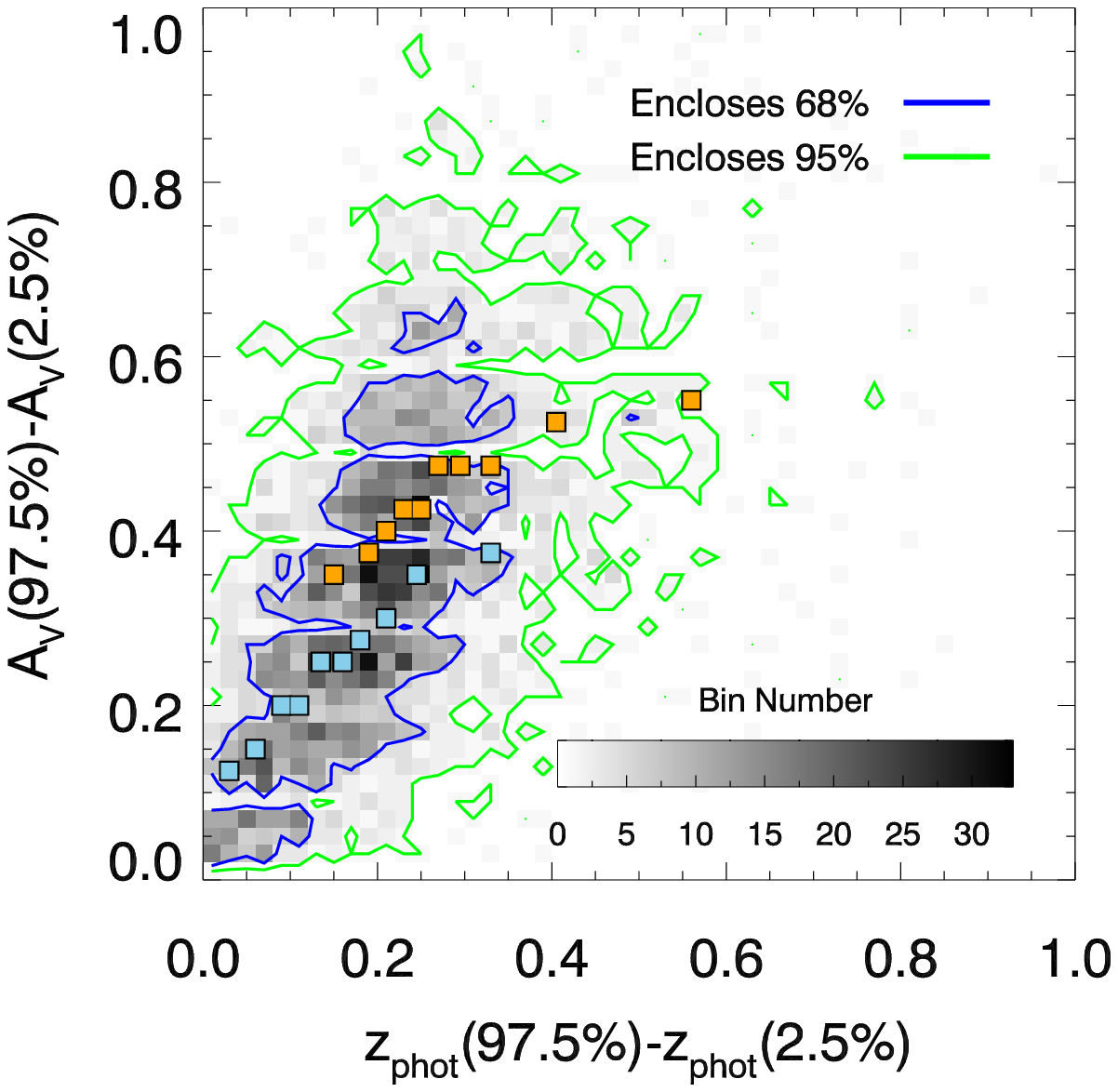} \\
\end{array}$
\end{center}
\vspace{-0.3cm}
\caption{Comparison between the photometric redshift uncertainty ($\pm$2$\sigma$ confidence range) and the uncertainty of $M_*$, SFR, $L_\mathrm{dust}$, and $A_V$ as derived from \magphysz. As expected, larger photo-$z$ uncertainties result in larger parameter uncertainties and are important to account for in studies relying on photometric redshifts to study galaxy evolution. The clumpiness of the $A_V$ panel is an artifact of the discretization of this parameter in the code. 
\label{fig:photz_err_vs_prop_err}}
\end{figure*}

A key feature of \magphysz\ is the self-consistent incorporation of photometric redshift uncertainty into the uncertainty of all derived properties. As a demonstration of the importance of this effect, we show the relation between the photometric redshift uncertainty ($\pm$2$\sigma$ confidence range; difference between 97.5 and 2.5 percentile range) and the uncertainty in a few of the derived physical properties in Figure~\ref{fig:photz_err_vs_prop_err}. We use the $\pm$2$\sigma$ confidence range to highlight a larger range of parameter space, but the effective trends are identical for the $\pm$1$\sigma$ ranges. For clarity, we do not include a small fraction of cases where the difference between the $\pm$2$\sigma$ range is zero, which is an artifact of the limited number of models that are in the generated SED libraries (this is most prevalent when models at the extremes of our priors are preferred; see Figure~\ref{fig:bump_z_prior}). As expected, the uncertainty range for parameters increases dramatically as a result of increases in the \zphot\ uncertainty. By accounting for this effect, \magphysz\ provides more realistic uncertainties for physical properties. These uncertainties affect studies relying on photometric redshifts to study galaxy evolution, such as determining galaxy scaling relations and their scatter \citep[e.g.,][]{brinchmann04, tremonti04, speagle14, cappellari16, tacconi18}.

\section{Discussion}\label{discussion}
\subsection{\texttt{MAGPHYS+photo-}\texorpdfstring{\MakeLowercase{\textit{z}}}{z_math} Self-Consistency Test}\label{self_consistency}
As a self-consistency test of the code, we perform runs using mock galaxy observation based on \magphys\ generated SED models. Performing this test allows us to check for any influence that our priors might have on biasing the \zphot\ solutions. The mock observations are constructed using the stellar population and dust emission models generated by \magphysz\ for the G10 broadband filter set.

To construct our mock observations, we first randomly select a stellar emission model of input redshift, $z_\mathrm{in}$, in the range of $0.4<z_\mathrm{in}<6$. Next, a compatible dust emission model is selected by requiring that the fractional contribution to the total dust luminosity from the diffuse ISM ($f_\mu$) and birth cloud components ($1-f_\mu$) of both models agree within 0.01 (i.e., energy balance in each component matches) and also that the redshift of each model be within 0.01. We then assign flux uncertainties and offset that are based on the distribution of values for the broadband data in the G10 catalog. For each filter, we introduce a constant offset in the model flux that is drawn randomly from a Gaussian distribution centered on zero and with a $1\sigma$ of $\pm15\%$ of the flux value. These offset values are typical of those found between the observational data and the \magphysz\ model fits for the G10 sample and are chosen to result in a best-fit $\chi^2$ distribution that is roughly comparable to those found for the real observations ($\bar{\chi}^2_\mathrm{mock}=1.67$; $\sigma(\chi^2_\mathrm{mock})=0.73$). To simulate the detection rates for each filter, we include non-detections in the flux values that are drawn and assigned. Finally, flux error values are assigned to each filter by drawing randomly from the distribution of fractional error values that are observed in the G10 sample. Using the method described above, we construct a catalog of mock broadband observations for 2,000 simulated galaxies. 

\begin{figure}
\begin{center}
\includegraphics[width=0.45\textwidth,clip=true]{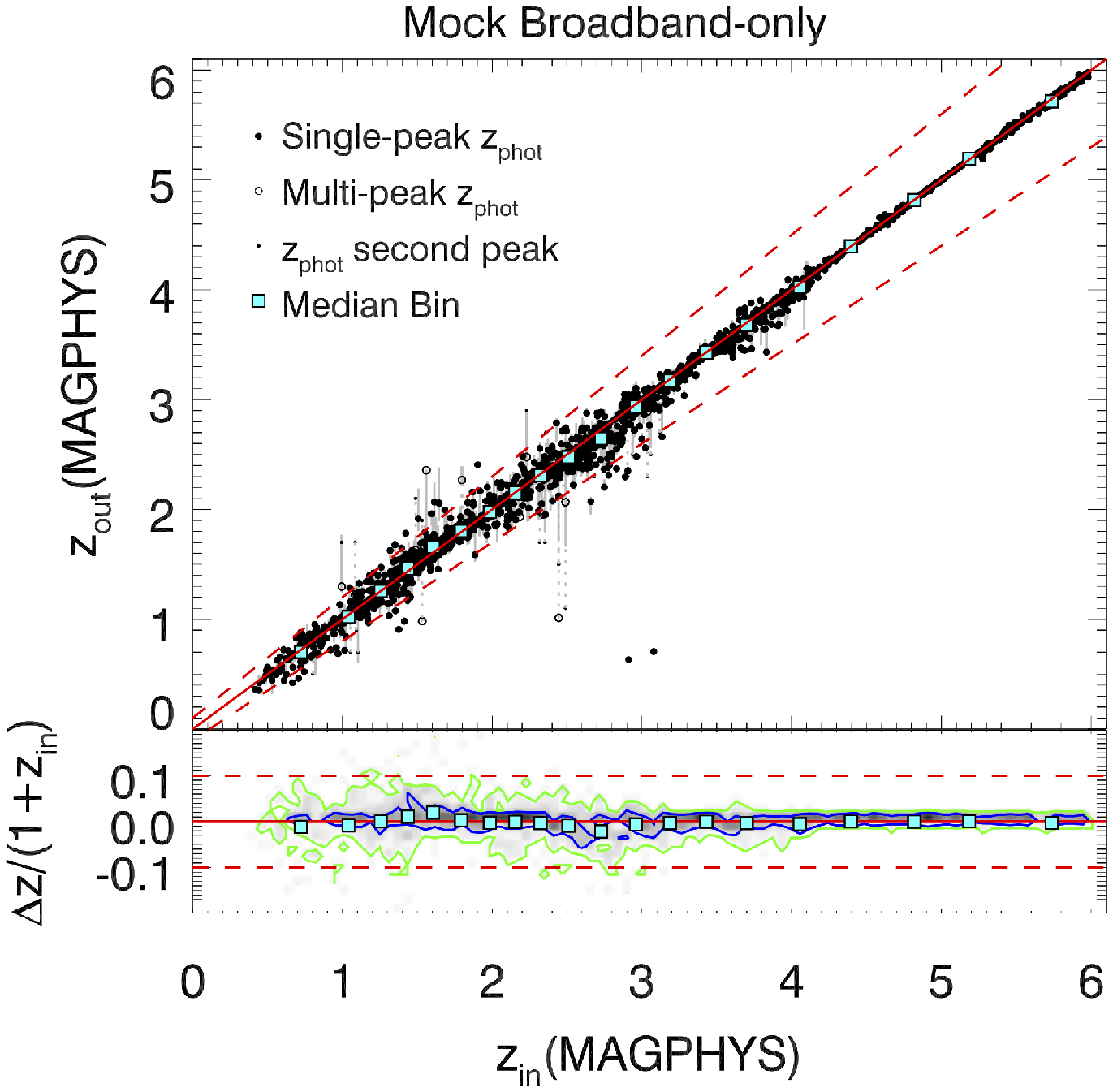}
\end{center}
\vspace{-0.4cm}
\caption{Comparison between the true redshift ($z_\mathrm{in}$) and photometric redshift ($z_\mathrm{out}$) for 2,000 mock galaxies using the G10 broadband filters of \magphys\ models (circles) that also include photometric offsets and uncertainties based on G10 catalog (see Section~\ref{self_consistency}). Labels are the same as in Figure~\ref{fig:photz_vs_specz}. The median values of equal-number bins are shown as cyan squares and show no significant systematic biases as a result of our parameter priors or adopted fitting method. The lower panel shows the redshift difference normalized by $1+z_\mathrm{in}$ in terms of surface density. 
\label{fig:photz_vs_specz_mock}}
\end{figure}

Figure~\ref{fig:photz_vs_specz_mock} shows the \magphysz\ output redshifts, $z_\mathrm{out}$, for the 2,000 mock SEDs. We note that for this test we do not incorporate any redshift dependence to the offset and fractional error values as we are most interested in seeing if there are systematic offsets at any redshift when the data are of comparable quality. For this reason, this test does not reproduce typical observational trends of increasing \zphot\ scatter at higher redshifts. It can be seen that there is excellent agreement over the entire redshift range from $0.4<z<6$ with minimal \zbias. The two instances of underestimated $z_\mathrm{out}$ at $z_\mathrm{in}\simeq3$ are the result of the Lyman break feature being incorrectly interpreted as a particularly strong 2175\ang feature in a very dusty mock galaxy. The better precision at the highest redshifts is the result of the Lyman break being very well sampled by the numerous optical filters at unrealistically high $S/N$. To test if higher uncertainty for the highest redshift mock galaxies makes a difference, we performed a separate test of models at $z>4$ with uncertainties increased by a factor of 30 from the previous run and still found no \zbias\ to be evident. These results demonstrate that the choice of priors, in particular the non-uniform prior for the model redshift distribution (Figure~\ref{fig:bump_z_prior}), and the fitting method adopted do not introduce significant \zbias\ in the results.

\subsection{Influence of IR Filters on photo-$z$ Estimates}
\begin{figure*}
\begin{center}
\includegraphics[width=1.\textwidth]{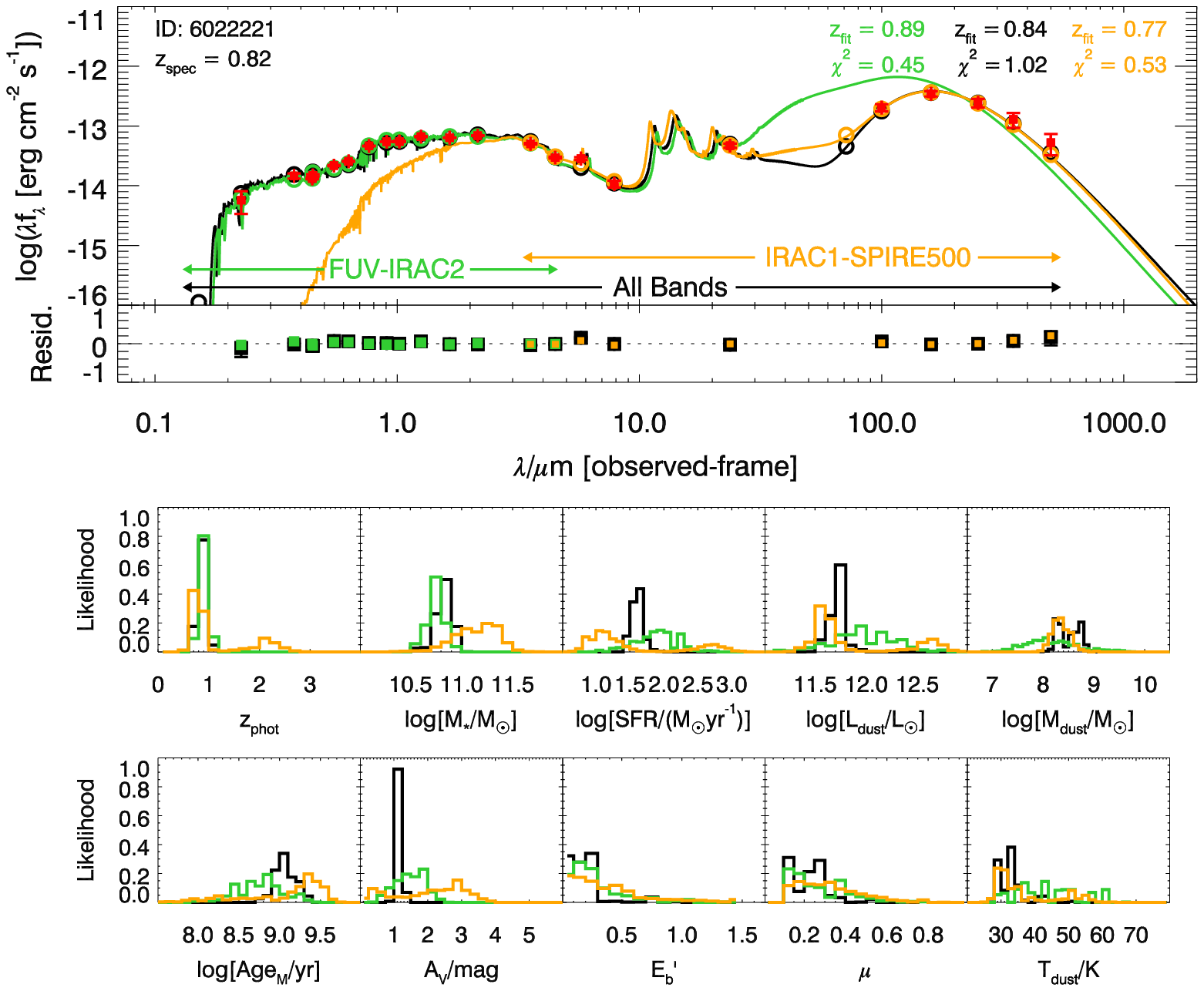}
\end{center}
\vspace{-0.1cm}
\caption{Similar to Figure~\ref{fig:magphys_example_COSMOS}, but demonstrating the \magphysz\ results for a randomly selected G10 galaxy when utilizing a subset of the available filters (colored curves) compared to the full filter set (black curve). The SEDs are shown in terms of flux instead of luminosity for easier comparison between the fits. The symbol sizes in the residual panel are varied for clarity. The fit using the FUV-IRAC2 bands (green curve) is able to constrain the redshift and stellar mass well, but does a poor job reproducing the other parameters. Conversely, the fit to the IRAC1-SPIRE500 bands provides a less precise redshift estimate and only provide constraints on the dust parameters. The IR-only fits often have multi-peak solutions for \zphot\ and other parameters due to a $z$-$T_\mathrm{dust}$ degeneracy. Using the full SED provides the tightest constraints on \zphot, and therefore also on the constraints for the other luminosity-dependent physical properties.  
\label{fig:filter_compare_magphys}}
\end{figure*}

Given that most photo-$z$ codes in the literature use only UV-NIR data, in this section we examine the influence that including the IR data have on constraining \zphot\ in \magphysz. In particular, we compare the fits obtained using only the UV through NIR filters (FUV-IRAC2; i.e., filters typically included in stand-alone photo-$z$ codes) and only the NIR through sub-mm filters (IRAC1-SPIRE500) relative to the fits when all of the available filters are used. Figure~\ref{fig:filter_compare_magphys} demonstrates the \magphysz\ solutions for a G10 galaxy when utilizing these different subsets of filters. In general we find that the UV-optical bands are crucial for precise estimation of \zphot\ whereas having IR-sub-mm bands alone are often not sufficient as \zphot\ tends to have a very broad likelihood distribution and/or have multiple peaks. This limitation is mainly due to the degeneracy in the location of the dust emission peak, because it varies both with $T_\mathrm{dust}$ and redshift. For example, when using only the IR-sub-mm bands in G10 galaxies with 5 or more bands at 24\micron\ or longer with $S/N>3$, we find 54\% of the median \zphot\ values (similar fraction for best-fit \zphot) are within $|\Delta z|/(1+z_\mathrm{spec})<0.1$, with 52\% of cases having multi-peak \zphot\ solutions. Thus, even for galaxies that sample the far-IR dust peak well, it is difficult to precisely constrain \zphot\ when $T_\mathrm{dust}$ is allowed to vary. However, if one has finer sampling of the PAH features, as will be possible in the future with \jwst, then the ability to constrain \zphot\ from IR-only bands will increase greatly. In addition, for very dusty sources, utilizing upper limits for UV-optical data in addition to IR detections will significantly improve the photometric redshift constraints because in many cases one of the ($A_V$-$z$) degenerate solutions can be ruled out \citep{daCunha15}. It is also worth noting that very dusty sources are also much easier to detect at observer-frame (sub-)mm than at other wavelengths owing to the negative $K$-correction approximately canceling the luminosity distance dimming at $z\gtrsim1$ \citep[e.g.,][]{blain&longair93}.

It is expected that the inclusion of the IR bands should increase the precision of \zphot\ in our samples, as having a wider wavelength sampling of the SED will always be beneficial and the IR luminosity also provides constraints on the dust attenuation. A comparison of the \zphot\ uncertainties for the G10 and C15+SD samples when using the full (broadband) filter set to those when only using the UV-NIR filters demonstrate that the former case has a $\pm$1$\sigma$ confidence range that is slightly lower than the latter, with a median($z$(84\%)-$z$(16\%))=0.075 vs. 0.105, respectively, for G10 and 0.125 vs. 0.155, respectively, for C15+SD. The changes in the values for $\sigma_\mathrm{NMAD}$, $\eta$, and \zbias\ are relatively minor between these runs. It is worth noting that the addition of IR-bands will have a larger impact on the \zphot\ precision in fields where there is sparser and/or shallower coverage at shorter wavelengths. The relatively minor improvement in precision afforded by the addition of IR-bands in the COSMOS field is primarily a testament to its extensive multiwavelength UV-NIR coverage and depth that provide exceptional SED sampling for \zphot\ determination. 

Following from the previous discussion, it is also worth briefly highlighting that users should be conscious of the fact that when there is little constraining information for a particular physical property, which can be due to gaps in spectral coverage, the returned posterior probability distribution (or confidence range) will be very broad (often reflecting the prior). As an example, one parameter that is often not well constrained is the total dust mass, $M_\mathrm{dust}$, as this requires sampling of the Rayleigh-Jeans tail of the dust graybody emission (rest-frame sub-mm wavelengths) because cold dust accounts for the bulk of the mass while warm dust dominates the total infrared luminosity \citep[e.g.,][]{scoville16}. Another example is the 2175\ang bump strength which will always have a positive value ($E_b'>0$) when there are no filters sampling that region because of the adopted prior distribution. It is important to account for the full posterior probability distribution when performing an analysis of any of the derived galaxy properties.

\subsection{Influence of the 2175\texorpdfstring{\ang}{Ang} Feature on Photometric Redshifts of Dusty Galaxies}\label{2175_zbias}
In this section we outline the effect that the 2175\ang absorption feature has on \zphot\ values when it is not accounted for in the galaxy models. It is worth noting that the influence of this feature on the attenuated SED scales directly with the amount of reddening ($A_{2175}=E_b\cdot E(B-V)\propto E_b'\cdot A_V$; this is ignoring effects of the filter response), such that the influence on \zphot\ is larger for dustier galaxies at a fixed value of $E_b'$. It is worth highlighting that because the bump strength parameter ($E_b'$) is independent of the total amount of dust reddening ($A_V$), there is a degeneracy in the manner that the total attenuation at the wavelength of the feature can be fit (i.e., low $E_b'$ and high $A_V$ vs. the opposite). This issue is mitigated when IR data is available to better constrain $A_V$ via energy balance. This also implies that galaxies with higher $A_V$ tend to have more reliable estimates for $E_b'$ because the total bump attenuation becomes more sensitive to small changes in $E_b'$.

\begin{figure}
\begin{center}
\includegraphics[width=0.45\textwidth]{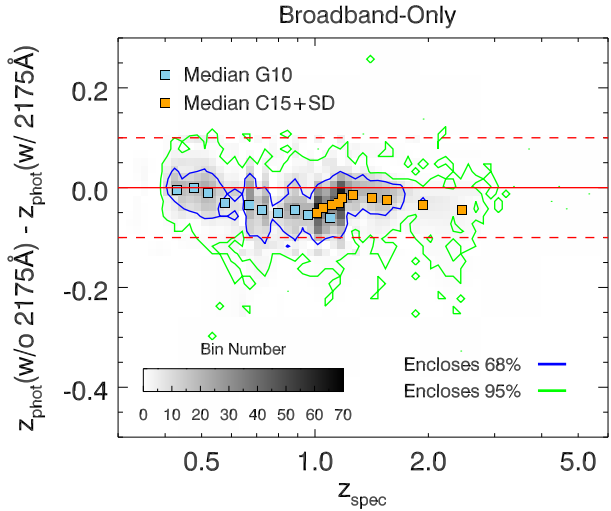}
\end{center}
\vspace{-0.5cm}
\caption{Difference between \zphot\ values obtained for our COSMOS samples when the additional 2175\ang absorption feature is not included in our models relative to when it is included. The \zphot\ values for runs without the 2175\ang feature are systematically lower at most redshifts, with a median offset of $z_\mathrm{phot}(\mathrm{w/o~2175\ang})-z_\mathrm{phot}(\mathrm{w/~2175\ang})=-0.03$.  The offset depends strongly on the number of filters that probe the 2175\ang region, their filter width, and their sensitivity ($S/N$). These effects result in the variation of the \zbias\ as a function of redshift. When the feature is not accounted for in our models, we systematically underestimate the correct redshift.
\label{fig:2175_zbias}}
\end{figure}

We run \magphysz\ without the prescription for the 2175\ang feature but keep every aspect the same (i.e., here we fix $E_b'=0$; see Section~\ref{magphys_method}) and compare them to our previous results. In Figure~\ref{fig:2175_zbias}, we show the difference in the \zphot\ values obtained for our samples as a function of \zspec\ for runs without a bump relative to those when a bump is allowed. It can be seen that the \zphot\ values for runs without the 2175\ang feature are systematically lower at most redshifts relative to those for which it is included. The offset depends strongly on the number of filters that probe the 2175\ang region, their FWHM, and their $S/N$. For example, the $u$-band is the only broadband filter probing the 2175\ang region at $0.4\lesssim z\lesssim0.6$, although weakly because the bump peak shifts to its effective wavelength at $z\simeq0.7$, and it typically has a lower $S/N$ than the $B$-band \citep[$m_\mathrm{AB}(3\sigma)=26.6$~mag and $27.0$ for 3" diameter aperture, respectively;][]{laigle16} such that there is a small offset at $0.4\lesssim z\lesssim0.6$ for this filter set. This is because higher weight is given to the $B$-band, together with the other optical bands, in the fits, which begins to probe the feature at $z\gtrsim0.6$ along with the $u$-band. The \zphot\ offset due to this feature in dusty galaxies can be larger if a filter probes this region and a few other bands surrounding the feature are available for fitting. The median redshift offset of the sample for runs without a 2175\ang feature is $z_\mathrm{phot}(\mathrm{w/o~2175\ang})-z_\mathrm{phot}(\mathrm{w/~2175\ang})=-0.03$ and this translates to an underestimate of the spectroscopic redshift, with $z_\mathrm{phot}(\mathrm{w/o~2175\ang})-z_\mathrm{spec}\simeq-0.03$ (not shown). This effect highlights that not properly accounting for this feature will result in a \zbias\ for dusty galaxies. In some photo-$z$ codes, a subtle \zbias\ can typically be corrected for by applying zeropoint offsets or template corrections in order to minimize model residuals at the spectroscopic redshift. This procedure is not performed in \magphysz\ because it would impact the ability to constrain physical properties by effectively altering the shape of the observed SED. We note that the 2175\ang feature has very little influence on the resulting physical property constraints, due to the minimal role it plays in the total attenuation energy budget, but that a \zbias\ effect will directly translate to biases in all luminosity-dependent quantities.

For all previous analyses, we have assumed characteristics for the shape of the 2175\ang feature based on the MW extinction curve. However, the assumption of a fixed central wavelength and FWHM may not be accurate for galaxy attenuation curves \citep[e.g.,][]{noll09a}. From fixed-redshift runs of standard \magphys, we find that the central wavelength of the absorption feature, based on intermediate band data, is very close to the MW value (Battisti et al. in prep.), suggesting that this central wavelength is a reasonable assumption. We also test the effect of reducing the width of the 2175\ang feature in the code to 300\AA, corresponding to the average value found by \citet{noll09a} for spectroscopically observed $z\simeq2$ galaxies, and find that this has a small effect on the \zphot\ results (typically $|\Delta z_\mathrm{phot})|\lesssim0.01$, which is below the scatter) and does not lead to significant improvement in the quality of the fits (based on $\chi^2$). We conclude that the \magphysz\ fitting results are not overly sensitive to the characterization adopted for the 2175\ang feature shape, especially when most general application will be on broadband data, but that accounting for additional absorption in this region is a required ingredient in the SED models to mitigate \zbias\ effects.

\subsection{Comparison with LePhare}\label{LePhare_comparison}
As a comparative test of \magphysz, we examine the photo-$z$ values in the COSMOS2015 catalog that are computed using the \lephare\ code \citep{arnouts02, ilbert06} as described in \citet{laigle16}. The \citet{laigle16} results are based on total flux values of all broad and intermediate-bands (3\arcsec + aperture correction) and these are the same values adopted in our analysis (Section~\ref{magphys_photoz}). As a brief summary, \lephare\ uses a set of 33 galaxy templates that include emission lines together with three extinction/attenuation curves considered: 1) the starburst attenuation curve \citep{calzetti00}, 2) a modified version of the starburst curve including a 2175\ang bump \citep{fitzpatrick&massa86}, 3) and the Small Magellanic Cloud extinction curve \citep{Prevot84}. An important aspect of \lephare\ that distinguishes it from \magphysz\ is that it computes systematic offsets that are applied to the predicted magnitudes in order to minimize the residuals with the observations for the spectroscopic sample \citep{ilbert06}. These offsets have the same effect as changing the photometric zeropoint of individual bands. In \magphysz\ we choose not to adopt a similar approach because it is equivalent to slightly altering the intrinsic shape of the observed SED and this impacts the derived physical properties.

\begin{figure*}
\begin{center}
$\begin{array}{cc}
\includegraphics[width=0.47\textwidth,clip=true]{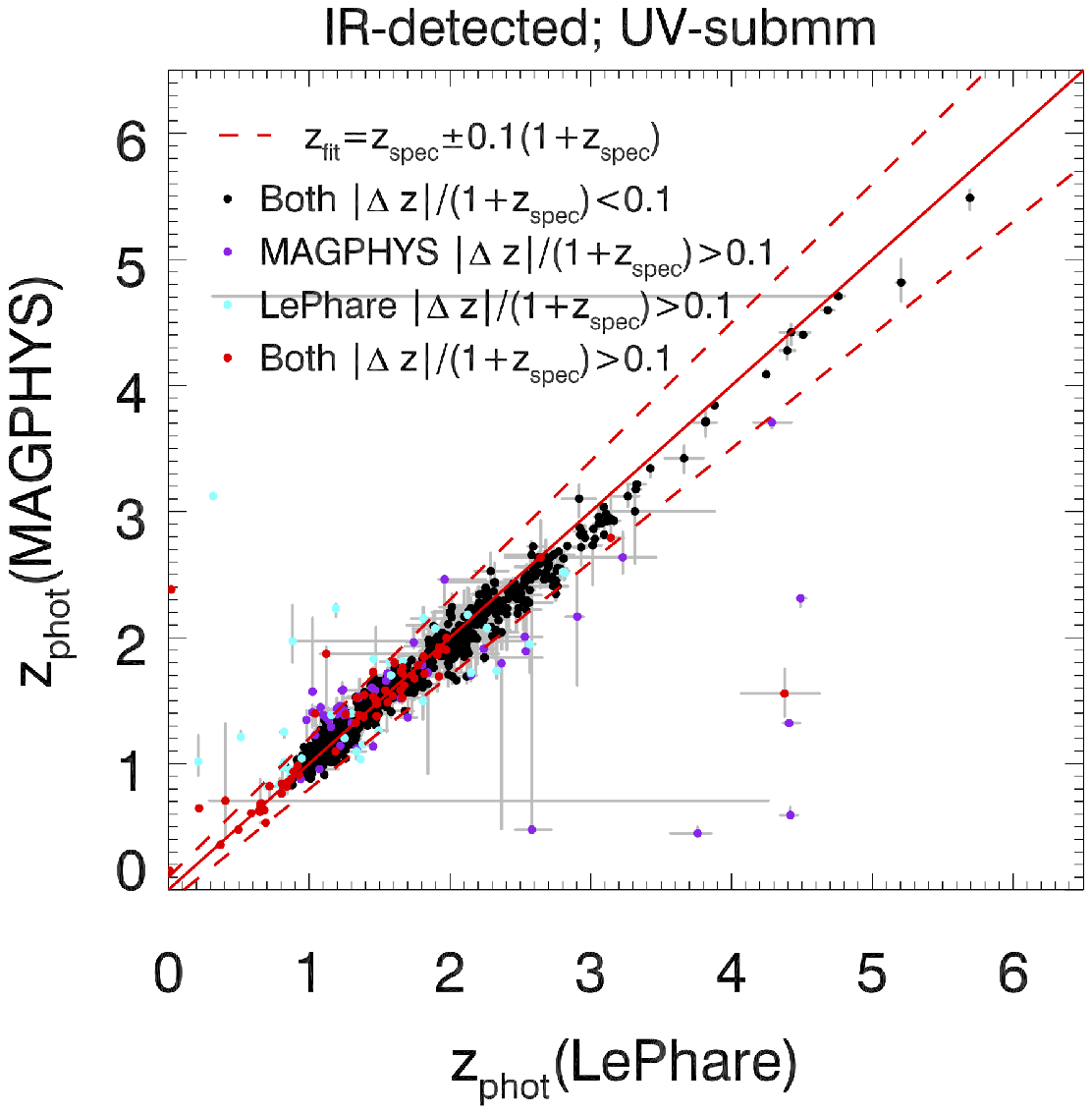} & \hspace{3mm}
\includegraphics[width=0.47\textwidth,clip=true]{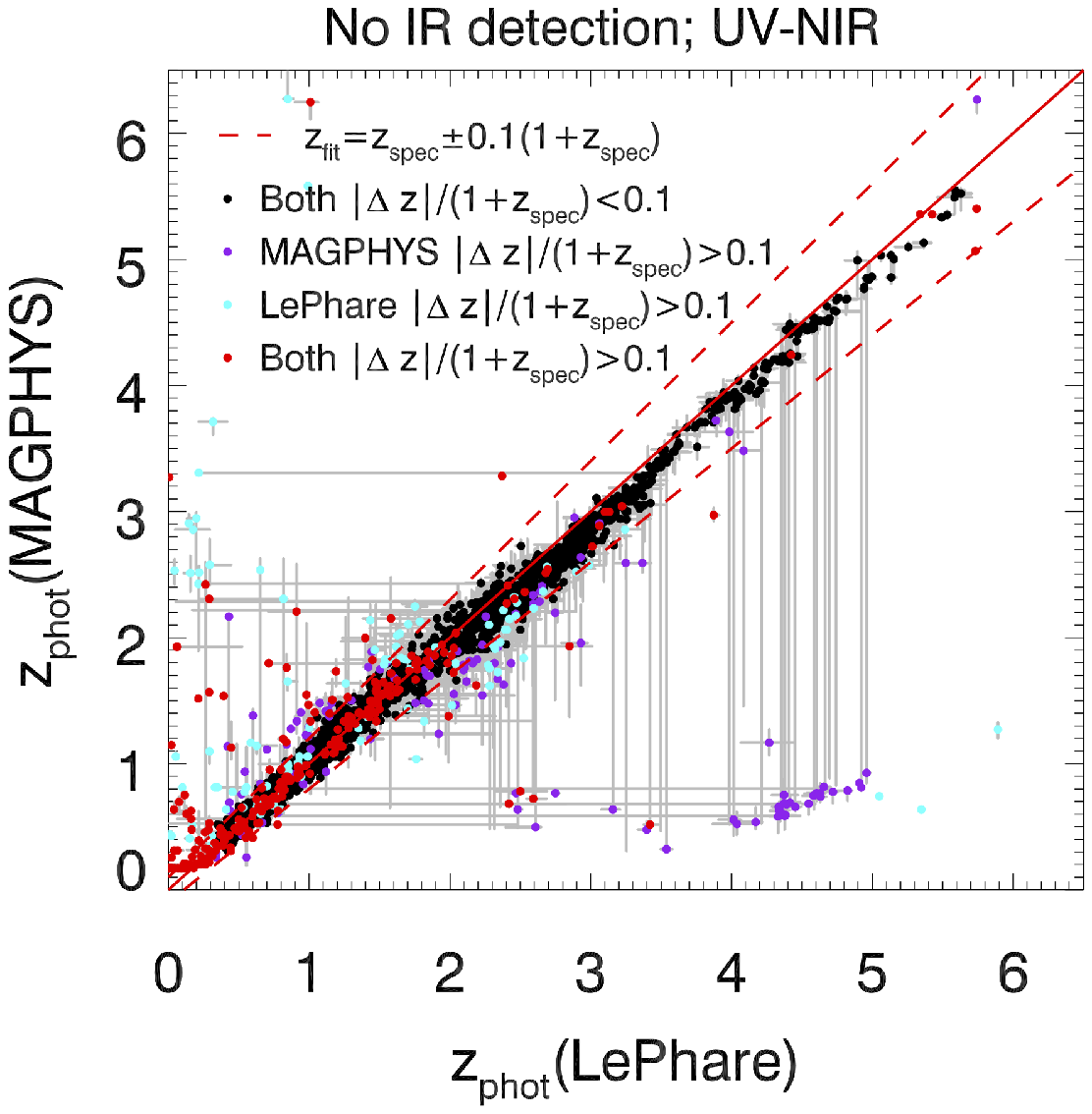} \\
\end{array}$
\end{center}
\vspace{-3mm}
\caption{Comparison between the photometric redshifts of \magphysz\ and \lephare\ for COSMOS2015 galaxies. (\textit{Left:}) Results for the IR-detected sample (C15+SD catalog). Both codes rely on the same photometric data, however \lephare\ uses only the observed UV-NIR data in the COSMOS2015 catalog for fitting. Black circles indicate cases where both codes provide the correct redshift within $|\Delta z|/(1+z_\mathrm{spec})<0.1$. Colored circles indicate cases where one code or both codes fail to estimate the correct redshift and have $|\Delta z|/(1+z_\mathrm{spec})>0.1$. (\textit{Right:}) Comparison for galaxies without an IR detection using the same photometric data (from NUV to IRAC ch4). There is a slight offset present in the values of \zphot\ at $z_\mathrm{spec}\gtrsim2$. We believe this to be the result of photometric effects (explored in Appendix~\ref{app_CANDELS}). In both cases, \magphysz\ performs very well with $\sim$95\% agreement with \lephare\ ($\sim$90\% both codes agree correctly, $\sim$5\% both agree incorrectly).
\label{fig:photz_vs_specz_lephare}}\vspace{4mm}
\end{figure*}

For our comparison, we use two subsamples of the COSMOS2015 catalog: 1) galaxies in our main IR-detected sample that requires at least one band $\geq$24\micron\ with $S/N>3$ and 2) galaxies without any IR detection (i.e., independent from the sample used in the previous analysis). Using a sample without the IR requirement enables a much larger sample of high-$z$ galaxies to be compared and will also span a wider range of galaxy properties. AGN are excluded in the IR sample using the methods described in Section~\ref{AGN_ID}, leaving 2,008 galaxies. We note that this is slightly larger than the previously used C15+SD sample because here we do not remove duplicates with the G10 sample. For the sample without IR detection, AGN are excluded using IRAC colors \citep{donley12}, radio \citep{seymour08}, and X-ray selection techniques. We also exclude sources with fewer than 8 bands detected at $S/N>3$ (removes $<1$\% of the sample; median number of detected filters is 26). To reduce the total computational cost of this sample, we randomly select 3,000 galaxies from the total population at $z<1.5$ but use all galaxies in the sample at $z\geq1.5$. These steps leave a final sample of 4,514 galaxies. 

We compare the output of the two codes for the IR-detected sample in Figure~\ref{fig:photz_vs_specz_lephare}, \textit{Left}. Both codes show agreement in their photometric values and the correct redshift ($|\Delta z|/(1+z_\mathrm{spec})<0.1$) for 92.6\% of the sample. We find that in 1.8\% of cases \magphysz\ has $|\Delta z|/(1+z_\mathrm{spec})<0.1$ and \lephare\ has $|\Delta z|/(1+z_\mathrm{spec})>0.1$ (i.e., \magphysz\ performs better), 2.4\% of cases where the opposite is true, and 3.1\% of cases where both cases have $|\Delta z|/(1+z_\mathrm{spec})>0.1$ (i.e., both codes obtain incorrect redshift). A comparison of photo-$z$ metrics for each code are shown in Table~\ref{tab:magphysz_lephare_compare} and are of comparable quality. There is a slight offset evident at $z\gtrsim2$ between \magphysz\ and \lephare\ \zphot\ values. This effect will be discussed further after comparing the sample without IR detections. The fraction of cases for which the \zphot\ 1$\sigma$ confidence ranges are in agreement with \zspec\ is 44\% and 35\% for \magphysz\ and \lephare , respectively.

We compare the output of the two codes for the sample without an IR detection in Figure~\ref{fig:photz_vs_specz_lephare}, \textit{Right}. For the following discussion, we are excluding 21 galaxies for which no \zphot\ is reported from \lephare. The codes show agreement ($|\Delta z|/(1+z_\mathrm{spec})<0.1$) in their photometric values for 89.5\% of the sample (black circles). We find that in 2.3\% of cases \magphysz\ produces redshifts with $|\Delta z|/(1+z_\mathrm{spec})<0.1$ and \lephare\ does not, 2.8\% of cases where the opposite is true, and 5.4\% of cases where both are below this accuracy. The larger disagreement in this sample by both codes is mostly due to these sources being fainter (lower $S/N$) than the IR selected sample. The comparison of photo-$z$ metrics are shown in Table~\ref{tab:magphysz_lephare_compare} and are similar between the codes. The fraction of cases for which the \zphot\ 1$\sigma$ confidence ranges are in agreement with \zspec\ is 41\% and 43\% for \magphysz\ and \lephare , respectively. It is worth noting that despite imposing a non-zero prior for the 2175\ang bump strength, the resulting photo-$z$ estimates for the non-IR sample show similar agreement as for the IR-detected sample.  In general, \lephare\ obtains slightly better \zphot\ metrics than \magphysz\ because it implements zeropoint offsets (i.e., `tunes' the data/models), it does not impose a minimum photometric uncertainty, and also includes emission lines. 

Looking at both panels of Figure~\ref{fig:photz_vs_specz_lephare}, there is a slight disagreement in \zphot\ values at $z\gtrsim2$ between \magphysz\ and \lephare. For the sample without IR detection, we find that \magphysz\ has $z\text{-}\mathrm{bias}(z>2)=-0.019$, whereas \lephare\ has $z\text{-}\mathrm{bias}(z>2)=0.009$. This indicates that \magphysz\ is slightly underestimating the value of \zphot\ for these samples at high redshifts compared to \lephare, which slightly overestimates the \zphot . We note that the updated \zphot\ values from \citet{davidzon17}, which used \lephare\ in a manner optimized for $z>2.5$ galaxies (see paper for details), show a lower offset for the higher redshift galaxies with a $z\text{-}\mathrm{bias}(z>2)=0.001$.  The \magphysz\ offset is not particularly large, but we sought to investigate the factors that could lead to an offset. The results from our self-consistency tests (Section~\ref{self_consistency}) would suggest that this offset is not a result of the methodology and instead may be attributed to either a shortcoming of our current models in representing the data or the result of a photometric zeropoint/measurement issues. One difference between \magphysz\ and \lephare\ is the attenuation prescription. However, we found that adopting a fixed attenuation curve of the diffuse ISM ($\propto\lambda^{-0.7}$), which is similar in shape to the starburst attenuation curve used in \lephare, did not lead to noticeable change the to average values of \zphot. A second difference between the codes is that \lephare\ modifies the zeropoint flux values to minimize $\chi^2$ values based on calibration with the \zspec\ sample and this is not performed in \magphys. We indirectly explore the effect of differences in the photometric dataset in Appendix~\ref{app_CANDELS} where we use independent photometric catalogs from CANDELS for high redshift galaxies. Those results do not show a significant \zbias\ at $z>2$ and indicates that the \zphot\ offset may be attributed to small zeropoint changes and/or differences in methodology. In rare instances, the lack of emission lines in our models, but which are included in \lephare , could also lead to differences in the \zphot\ estimates. We plan to revisit the issue of a possible \zbias\ in \magphysz\ as it undergoes continued testing on a broader range of independent photometric datasets in the future. However, in its current state, it appears to perform well and achieves similar levels of accuracy as ``classic'' photo-$z$ codes.

\begin{table}
\begin{center}
\caption{Comparison of photo-$z$ metrics (defined in Section~\ref{broad_photoz}) between \magphysz\ and \lephare.}
\begin{tabular}{lccc}
\hline
 & \multicolumn{3}{c}{IR-detected Sample} \\
Code & $\sigma_\mathrm{NMAD}$ & $\eta$ & $z\text{-}\mathrm{bias}$ \\
\hline
\magphysz  & 0.027 & 0.033 & -0.010 \\
\lephare      & 0.022 & 0.029 & -0.005 \\
\hline \\
\hline
 & \multicolumn{3}{c}{No IR detection Sample} \\
Code  & $\sigma_\mathrm{NMAD}$ & $\eta$ & $z\text{-}\mathrm{bias}$ \\
\hline
\magphysz  & 0.031 & 0.055 & -0.010 \\
\lephare      & 0.018 & 0.056 & -0.002 \\
\hline
\end{tabular}
\label{tab:magphysz_lephare_compare}
\end{center}
\end{table}

In addition to \zphot, \lephare\ also estimates the stellar mass ($M_*$) of each galaxy using a library of synthetic spectra generated with the stellar population models of \citet{bruzual&charlot03} and these are also included in the COSMOS2015 catalog \citep[for details, see][]{laigle16}. The same stellar population models are also utilized in \magphys, along with the same initial mass function \citep{chabrier03}, and thus differences in the $M_*$ between the codes (for galaxies at the same \zphot) should mainly be attributed to different treatments/assumptions for the SFH, metallicity, and/or dust attenuation. The $M_*$ is generally the most reliable physical parameter constrained by SED modeling, typically varying by $\sim0.3$~dex or less \citep[e.g.,][]{conroy13, hunt19}. This is because the mass-to-light ratios in the optical-NIR regions (redward of $V$-band) are relatively stable across a wide range of SFHs and other assumptions \citep[e.g.,][]{bell&dejong01, zibetti09}.

\begin{figure*}
\begin{center}
$\begin{array}{cc}
\includegraphics[width=0.45\textwidth,clip=true]{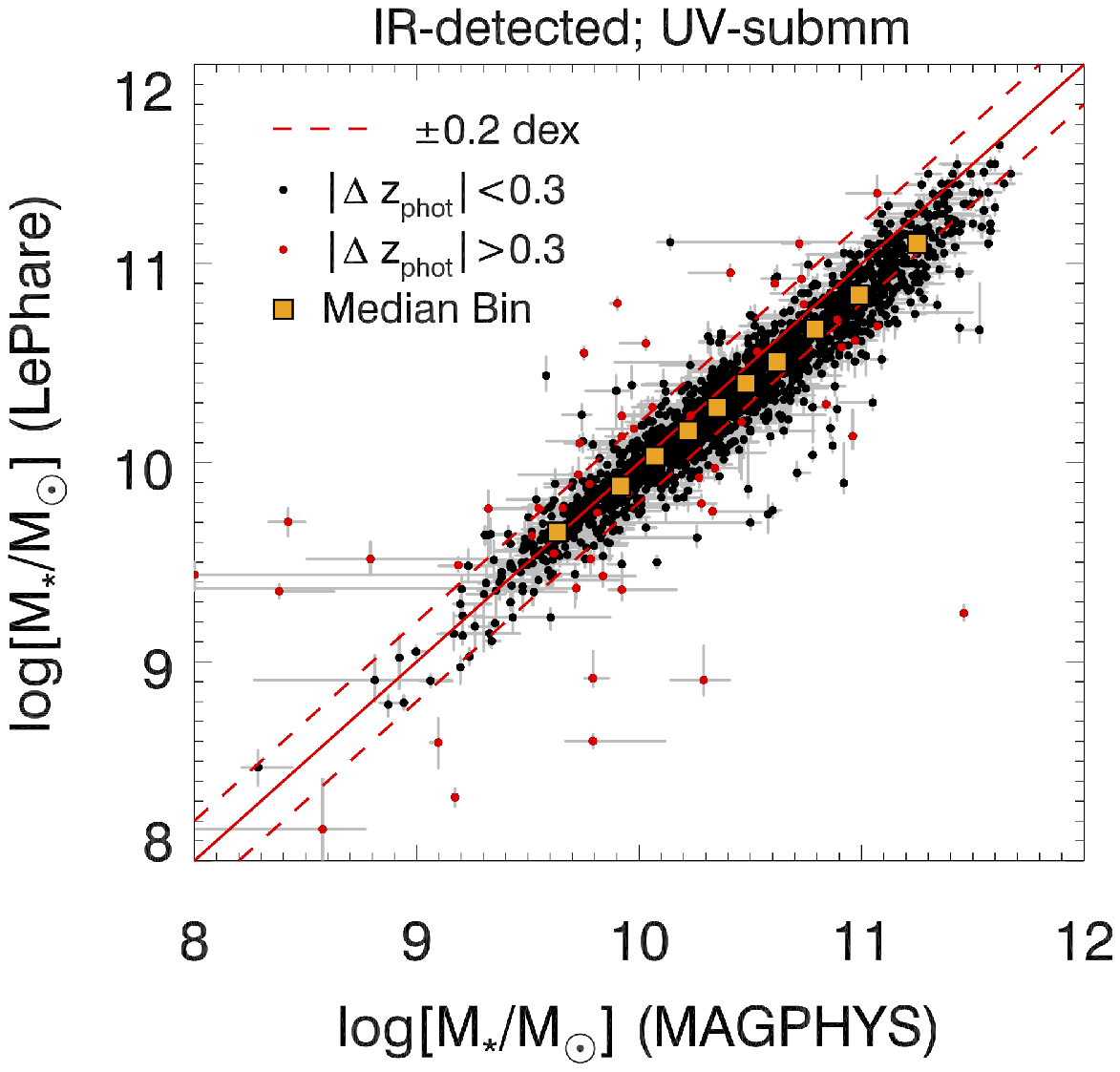} &
\includegraphics[width=0.45\textwidth,clip=true]{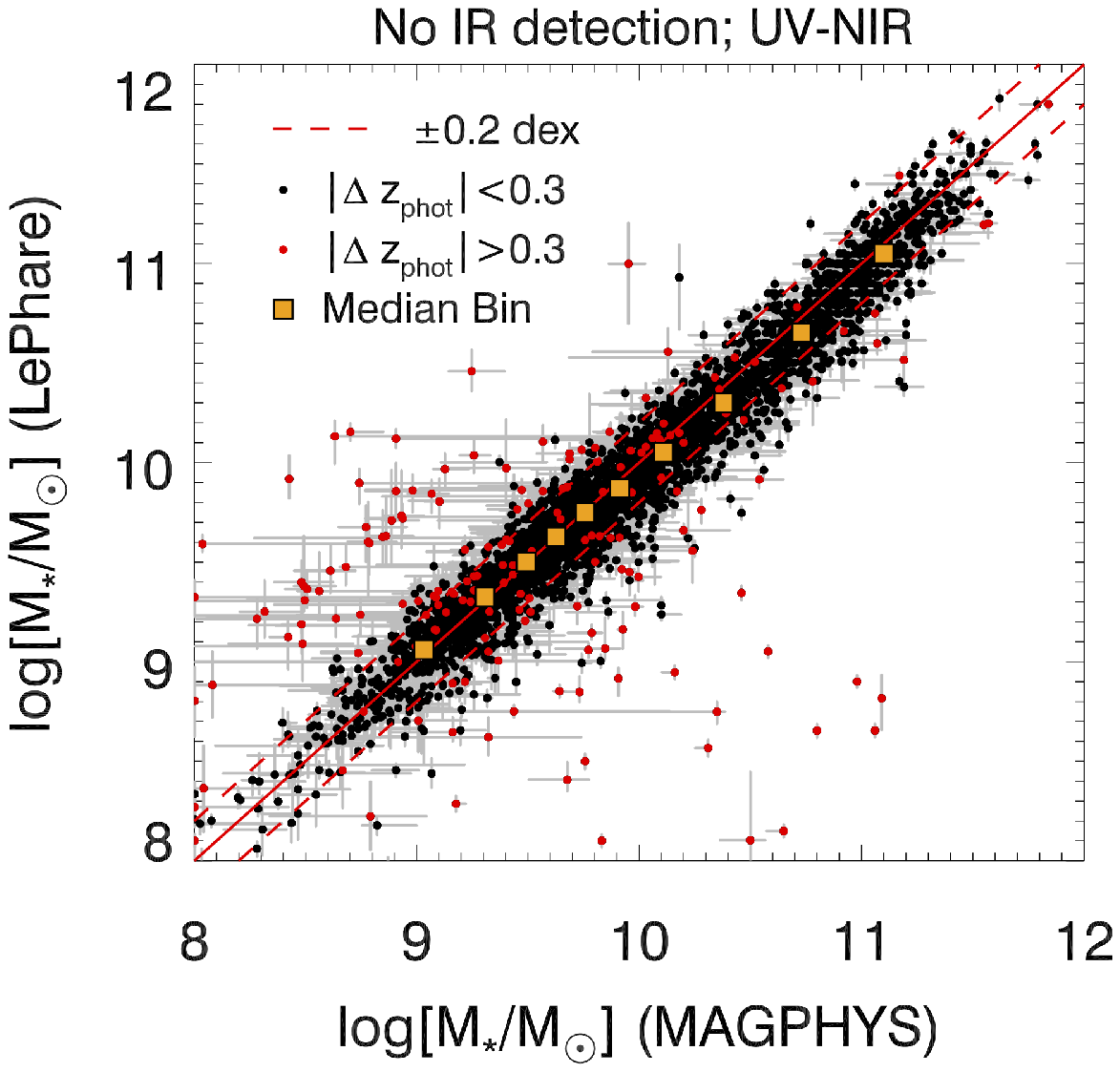} \\ 
\includegraphics[width=0.45\textwidth,clip=true]{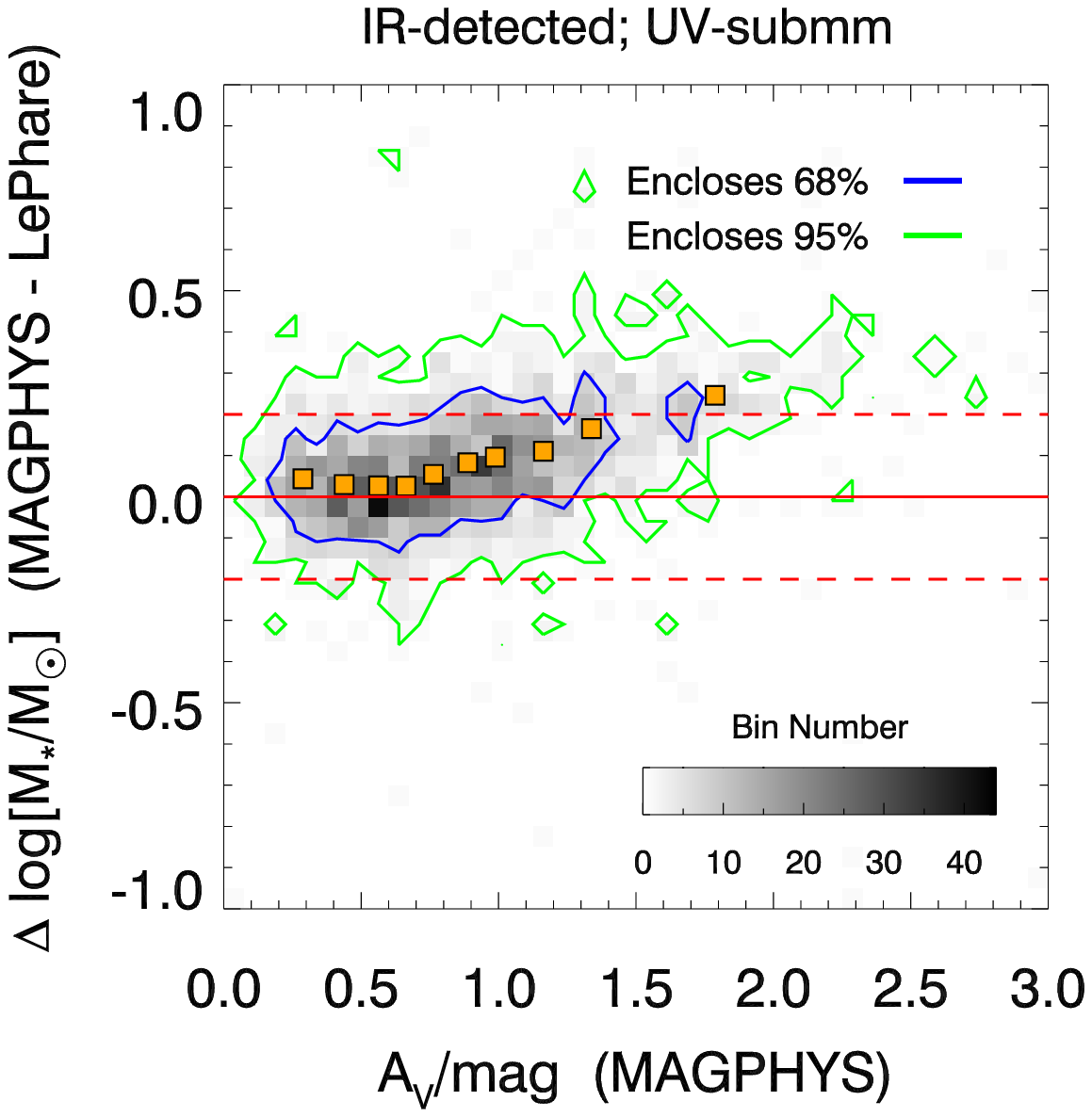} &
\includegraphics[width=0.45\textwidth,clip=true]{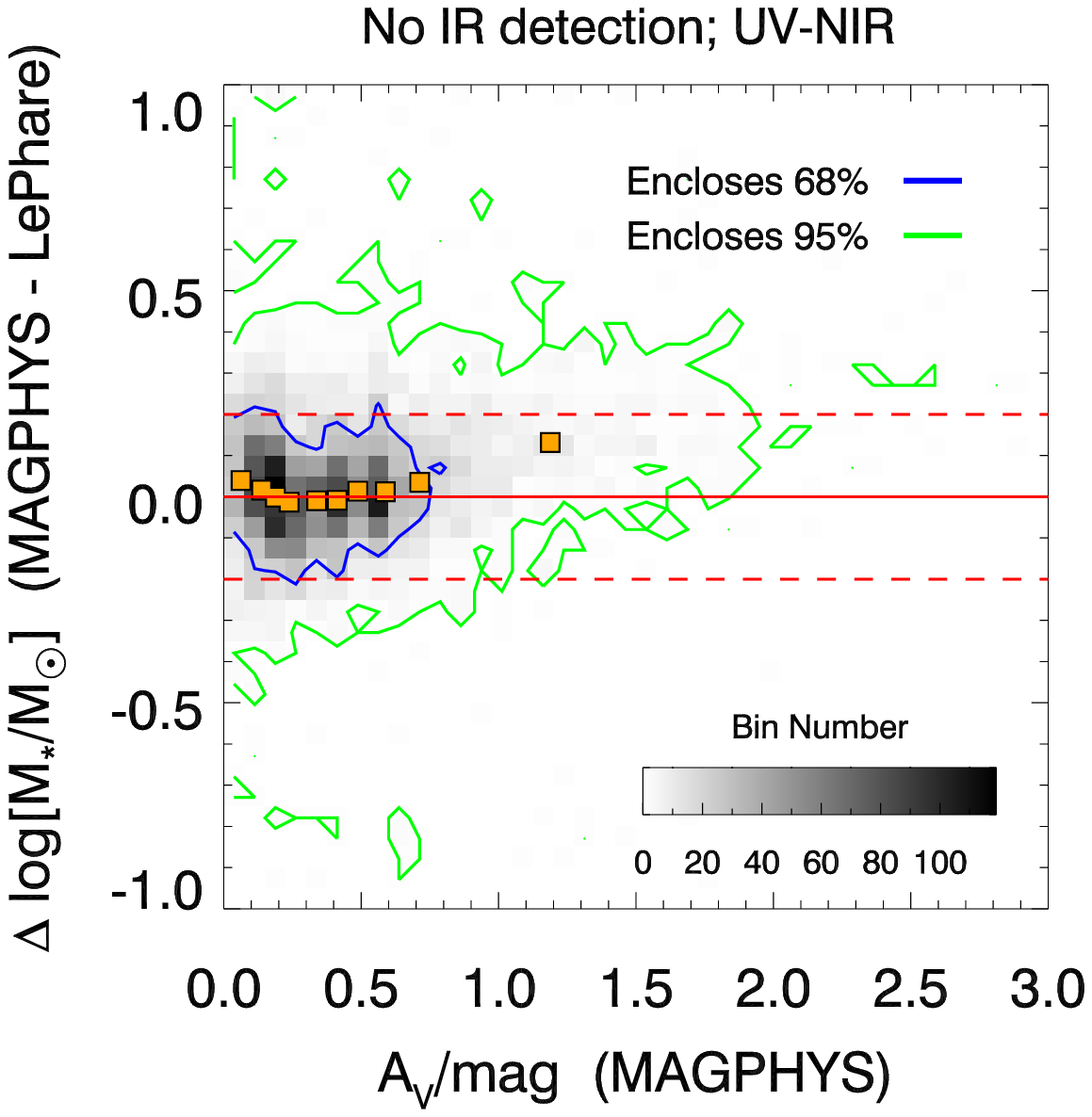} \\
\end{array}$
\end{center}
\vspace{-0.3cm}
\caption{Comparison between the stellar mass ($M_*$) inferred from \magphysz\ and \lephare\ for COSMOS2015 galaxies. \textit{Top:} Direct stellar mass comparison for the IR-detected sample (\textit{Left}) and the sample without any IR detection (\textit{Right}). Black circles indicate cases where the \zphot\ of both codes agree within $|\Delta z_\mathrm{phot}|<0.3$. Red circles indicate cases where \zphot\ disagree. The two codes show a sizable disagreement for the high-mass, IR-detected galaxies, which is related to the inferred amount of dust attenuation (see lower panels). In contrast, the non-IR sample show much closer agreement because of the lower levels of attenuation. \textit{Bottom:} Difference in stellar mass as a function of the effective $V$-band dust attenuation ($A_V$) for each sample. There is a significant trend where dustier galaxies have systematically higher stellar masses in \magphysz\ relative to \lephare, with the highest $A_V$ cases differing by up to a factor of two ($\sim$0.3~dex). This effect is primarily a consequence of inferring significant dust attenuation on the SED at optical-NIR wavelengths (e.g., compare intrinsic and attenuated SEDs in Figure~\ref{fig:magphys_example_COSMOS}).
\label{fig:mass_compare_lephare}}
\end{figure*}

We compare the $M_*$ inferred from \magphysz\ and \lephare\ for the two COSMOS2015 subsamples in Figure~\ref{fig:mass_compare_lephare} (\textit{Top}). It can be seen that for the IR-detected sample, which is biased toward dustier galaxies, the inferred $M_*$ from \magphysz\ is systematically higher than \lephare, being most disparate at the highest masses ($\gtrsim10^{10}$~$M_\odot$). We examine the influence of the assumed dust attenuation by comparing the difference in $M_*$ as a function of $A_V$ in Figure~\ref{fig:mass_compare_lephare} (\textit{Bottom}). There is a significant trend where larger differences occur for dustier galaxies (these also tend to be more massive), with the highest $A_V$ cases differing in $M_*$ by up to a factor of two ($\sim$0.3~dex). The change in $M_*$ with increasing $A_V$ is primarily a consequence of significant levels of dust attenuation on the SEDs in the optical-NIR region for the dustiest galaxies, which is required through the assumption of energy balance. Without the constraints from the IR data, the dust attenuation for these sources is underestimated in \lephare. In fact, 53\% of the IR-detected sample have $E(B-V)=0$ ($A_V=0$) for the best-fit \citet{bruzual&charlot03} model in \lephare\ (\texttt{extinction} parameter in COSMOS2015 catalog). An example of significant optical-NIR attenuation for a massive, dusty galaxy can be seen in Figure~\ref{fig:magphys_example_COSMOS} by comparing the best-fit intrinsic and attenuated model SEDs in the upper panel. However, it is worth emphasizing again that the IR-detected sample is highly biased toward cases where codes only relying on UV-NIR data, such as \lephare, are prone to have issues. Indeed, when we examine the sample of galaxies without IR detection (typically less dusty galaxies), the $M_*$ values are in much closer agreement. Looking at the difference in $M_*$ as a function of $A_V$ for those cases, it can be seen that the agreement coincides with the low inferred values for $A_V$ in this sample. However, similar to before, the trend of large mass difference is seen for the small fraction of cases with large inferred $A_V$. For reference, the 1$\sigma$ dispersion in the mass difference for the different $A_V$ bins in both samples are $\sim$30\% ($\sim$0.12~dex). These results suggest that the additional IR coverage provided in \magphysz\ is very important for accurate stellar mass estimation of dusty galaxies.

\section{Conclusion}\label{conclusion}
We have developed an extension of the spectral modeling code \magphys, called \magphysz , that allows the estimation of galaxy redshift simultaneously with other physical properties and their uncertainties in a manner that incorporates the redshift uncertainty. The success of the code in estimating photometric redshifts and physical properties is demonstrated for galaxies at $0.4<z<6.0$ in the COSMOS field. The main results of the paper are summarized below:
\begin{itemize} 
\item For IR-detected COSMOS galaxies, we achieve high photo-$z$ precision ($\sigma_\mathrm{NMAD}\lesssim0.04$), high accuracy (i.e., minimal offset biases; $\mathrm{median}(\Delta z/(1+z_\mathrm{spec}))\lesssim0.02$), and low catastrophic failure rates ($\eta\simeq4\%$) over all redshifts. These results are comparable to those obtained with the \lephare\ code.
\item The derived physical properties of galaxies using \magphysz\ are consistent with those derived from the fixed-redshift (\zspec) using standard \magphys.
\item We demonstrate a strong correlation between photometric redshift uncertainty and the uncertainty of other derived physical properties that is critical to account for in studies of galaxy evolution based on photo-$z$'s.
\item The inclusion of a weak 2175\ang absorption feature in the attenuation curve model is required to remove a subtle systematic \zphot\ offset ($z_\mathrm{phot}-z_\mathrm{spec}\simeq-0.03$) that would otherwise be present for our sample of IR-detected galaxies. 
\item We compare stellar masses inferred from \magphysz\ and \lephare\ and find that for dusty star-forming galaxies (large $A_V$) the $M_*$ can be underestimated by up to a factor of 2 ($\sim$0.3~dex) in the latter, being most disparate at the highest stellar masses ($\gtrsim10^{10}$~$M_\odot$). This highlights an issue that can arise for dusty galaxies in codes attempting to estimate properties from rest-frame UV-NIR data alone.
\end{itemize}
\magphysz\ is unique among existing codes in that it includes IR, sub-mm, and radio data when constraining \zphot. This is particularly beneficial for dusty, high-$z$ galaxies that are often undetected at rest-frame UV-optical wavelengths. The main benefit to utilizing \magphysz\ is that there is no need for users to run their catalogs through dedicated photo-$z$ codes prior to determination of galaxy physical properties. The self-consistent photo-$z$ and SED modeling capability of this code provides users with a reliable and easy method for interpreting large photometric datasets to study galaxy evolution.

\section*{Acknowledgments} The authors thank the anonymous referee whose suggestions helped to clarify and improve the content of this work.
EdC gratefully acknowledges the Australian Research Council as the recipient of a Future Fellowship (project FT150100079). Parts of this research were supported by the Australian Research Council Centre of Excellence for All Sky Astrophysics in 3 Dimensions (ASTRO 3D), through project number CE170100013. AJB thanks C. Laigle for correspondence regarding COSMOS2015 catalog usage. AJB also thanks 
S. Charlot, I. Davidzon, and O. Ilbert for comments that improved the manuscript. AJB is also thankful for attending ASTRO 3D writing retreats that provided a helpful environment to complete portions of this manuscript. We thank the invaluable labor of the maintenance and clerical staff at our institutions, whose contributions make our scientific discoveries a reality. Based on data products from observations made with ESO Telescopes at the La Silla Paranal Observatory under ESO programme ID 179.A-2005 and on data products produced by TERAPIX and the Cambridge Astronomy Survey Unit on behalf of the UltraVISTA consortium.

\bibliography{AJB_bib}

\clearpage

\appendix
\section{Reliability of photo-\texorpdfstring{\MakeLowercase{\textit{z}}}{z_math} for AGN}\label{app_AGN}
Here we briefly examine the photo-$z$ results for the AGN sources in our COSMOS samples using \magphysz . These include 179 and 117 AGN in the G10 and C15+SD samples, respectively. We distinguish this sample as IR, radio, and/or X-ray AGN, based on the diagnostic utilized to identify them. For G10, the breakdown is 126, 42, and 13, respectively. For C15+SD, the breakdown is 45, 64, and 26, respectively. Note that sources can be identified with multiple diagnostics (there are 2 and 9 multiply diagnosed AGN in the G10 and C15+SD samples, respectively). It is also worth stating that these selections are incomplete due to different depths for the various bands utilized. 

\begin{figure*}
\begin{center}
\includegraphics[width=0.55\textwidth,clip=true]{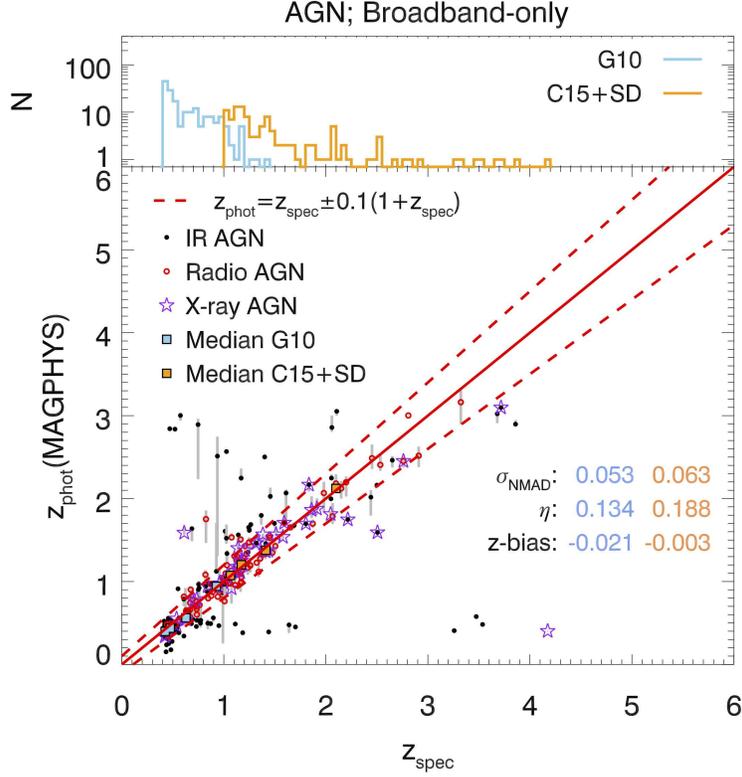}
\end{center}
\vspace{-0.2cm}
\caption{Comparison between the photometric and spectroscopic redshifts for AGN in the COSMOS field using \magphysz . The diagnostic used to identify the AGN is indicated as a solid black circle, open red circle, and open purple star symbols for IR, Radio, and X-ray methods, respectively. Note, some cases are identified by multiple methods. Despite lacking AGN models, the code still performs reasonably well in estimating \zphot\ for a majority of cases. The majority of catastrophic failures coincide with IR AGN with high $\chi^2$ values, indicating the AGN is significantly affecting the SED and not being fit well.
\label{fig:photz_vs_specz_AGN}}
\end{figure*}

As a reminder, current versions of \magphys\ do not include AGN models, although this is planned for an upcoming release, and therefore we expect the code to perform worse for these sources. We do not explore the quality of physical property estimates for AGN sources because this would require a comparison to fits that include AGN models and is beyond the scope of this paper. The \zphot\ results are shown in Figure~\ref{fig:photz_vs_specz_AGN} along with the corresponding quality metrics for each sample ($\sigma_\mathrm{NMAD}$, $\eta$, and $z\text{-}\mathrm{bias}$). Different symbols are used for the 3 AGN classes. These results indicate that despite the lack of AGN models, the \zphot\ estimates for AGN are still reasonably well determined in most cases, with only a slightly higher $\sigma_\mathrm{NMAD}$ than for the normal galaxy sample. However, the instance rate of catastrophic failures is noticeably larger, being almost 20\% for the C15+SD sample. 

The AGN class that appears to be most susceptible to catastrophic failure are those identified using IR diagnostics. The IR AGN make up 88\% and 91\% of the catastrophic failures for the G10 and C15+SD samples, respectively. This is not surprising since our SED fits include the IR regime but not the radio (for these runs) or X-ray. However, a significant fraction of the IR AGN still obtain reasonable \zphot\ estimates, with 83\% and 49\% of these cases having $\Delta z/(1+z_\mathrm{spec})<0.15$ for the G10 and C15+SD samples, respectively. The catastrophic failure cases are primarily instances where the AGN is dominating the IR SED and, as will be discussed next, this can be gauged by the $\chi^2$ values of the best-fit SED. In contrast, the radio AGN make up only 8.3\% and 4.5\% of the catastrophic failures for the G10 and C15+SD samples, respectively, and the X-ray AGN are only 4.2\% and 9.1\%, respectively. Over 90\% of the radio and X-ray AGN have $\Delta z/(1+z_\mathrm{spec})<0.15$ for both samples.

As expected, the \magphysz\ fits to the AGN sources also have considerably higher $\chi^2$ values on average than those for the normal galaxies. We find that 30\% and 42\% of the G10 and C15+SD AGN sample, respectively, lie above the $\chi^2>\bar{\chi}^2+4\sigma(\chi^2)$ thresholds discussed in Section~\ref{magphys_photoz}. We recommend users adopt a $\chi^2$ threshold cut to galaxy samples to reject possible AGN, stellar, or other sources that are not accounted for in the current models. The distribution of $\chi^2$ values will change from sample to sample, depending on the photometric quality and which regions of the SED are sampled. Therefore, it is best to make cuts based on the $\chi^2$ distribution for each sample instead of using a specific $\chi^2$ value across different samples. For example, imposing the cut $\chi^2>\bar{\chi}^2+4\sigma(\chi^2)$, discussed in Section~\ref{magphys_photoz}, would remove 71\% and 91\% of the AGN sources experiencing \zphot\ catastrophic failures in the G10 and C15+SD samples, respectively. The sources with the highest $\chi^2$ values correspond to cases with a dominant influence of the AGN on their IR SED and therefore they are more poorly constrained in \magphysz . Unfortunately, as the number of available bands being fit decreases, it becomes more difficult to identify AGN sources for exclusion using the $\chi^2$ metric.

\section{\texttt{MAGPHYS+photo-}\texorpdfstring{\MakeLowercase{\textit{z}}}{z_math} Results using CANDELS Photometry}\label{app_CANDELS}
In Section~\ref{LePhare_comparison}, we find evidence the \zphot\ values for COSMOS2015 that galaxies at $z\gtrsim2$ are slightly underestimated relative to \zspec. To examine if this is due to a potential issues with \magphysz , we explore if similar offsets are found for galaxies with independent photometry from CANDELS \citep{grogin11, koekemoer11}. We utilize the CANDELS photometric catalogs that are based on using the \texttt{Sextractor} version 2.8.6 \citep{bertin&arnouts96} in dual mode with the WFC3 F160W as the detection band \citep[for details see][]{nayyeri17}.

\begin{figure*}
\begin{center}
$\begin{array}{cc}
\includegraphics[width=0.47\textwidth,clip=true]{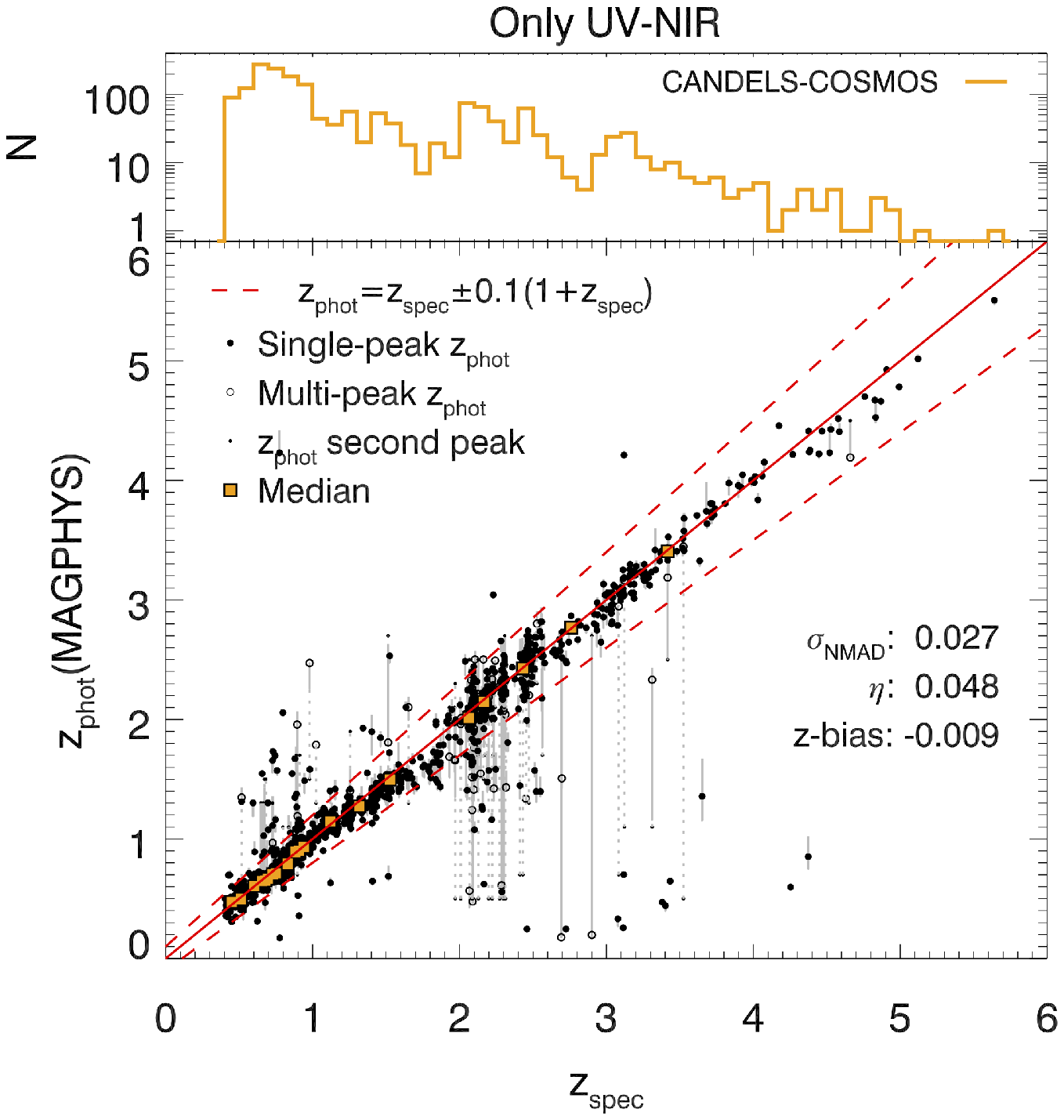} &
\includegraphics[width=0.47\textwidth,clip=true]{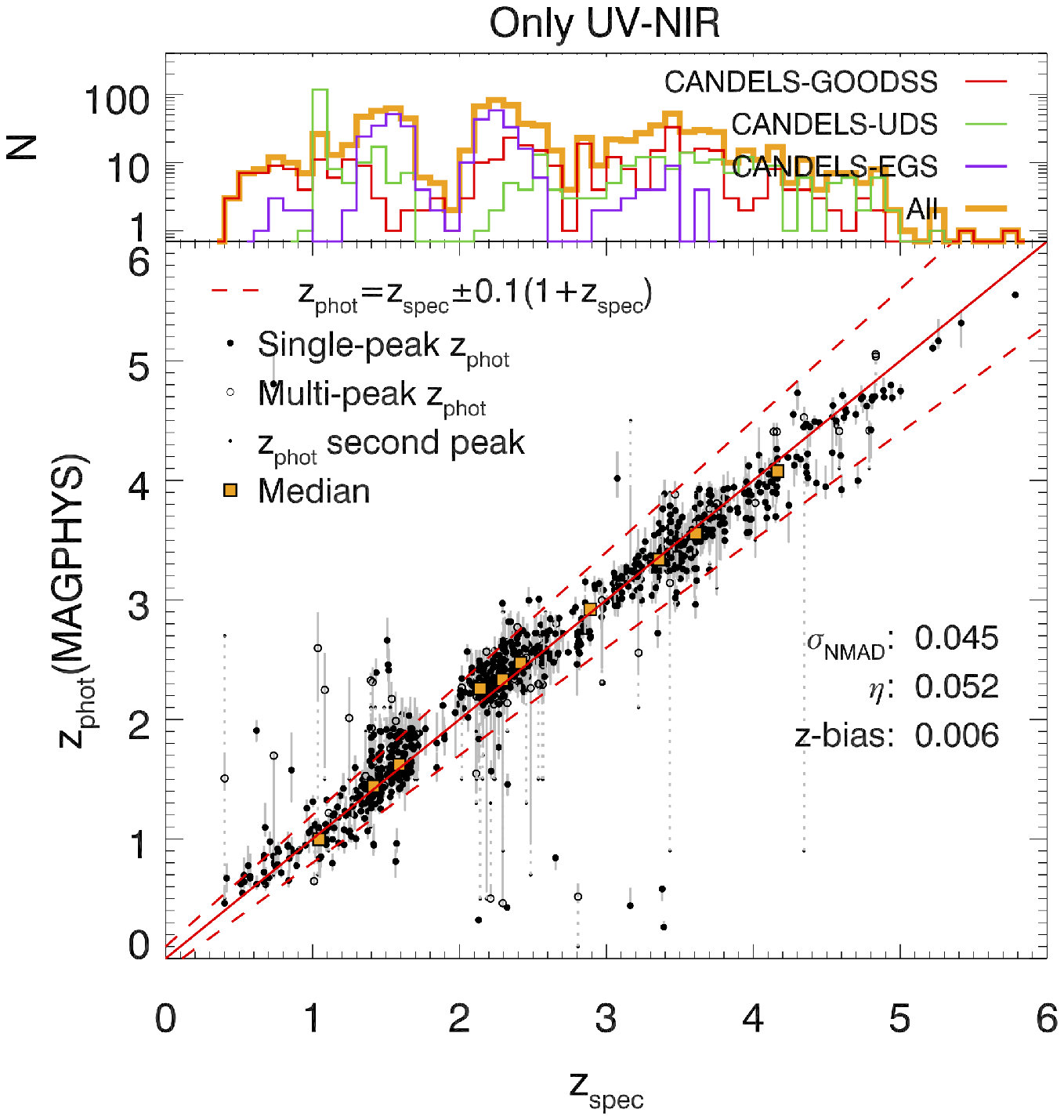} \\
\end{array}$
\end{center}
\vspace{-0.2cm}
\caption{\textit{Left:} Comparison between the photometric and spectroscopic redshifts for CANDELS-COSMOS galaxies using \magphysz . Labels are the same as in Figure~\ref{fig:photz_vs_specz}. No significant \zbias\ is found at $z>2$ using this dataset in contrast to the slight bias found for the non-IR COSMOS2015 sources. \textit{Right:} Comparison between the photometric and spectroscopic redshifts using CANDELS data in the GOODS-S, UDS, and EGS fields. The agreement is again quite good with minimal \zbias. These results highlight that photometric differences (either through zeropoints or methodology) can introduce small systematic offsets in the \zphot\ estimates.
\label{fig:photz_vs_specz_CANDELS}}
\end{figure*}

First, we use the catalog of the COSMOS field from \citet{nayyeri17}, which spans filters from $u$ to \spitzer/IRAC ch4, and cross-match those objects with the COSMOS master spectroscopic catalog (curated by M. Salvato). The photometric data include the same filters as the G10 and COSMOS2015 catalogs, as shown in Table~\ref{tab:filt_extinct} (up to IRAC ch4), with additional filters from \hst/ACS, \hst/WFC3, and Mayall/NEWFIRM. It is important to note that for this analysis the longest wavelength being utilized is IRAC ch4 (as is the case for the non-IR galaxies Section~\ref{LePhare_comparison}). These cross-matched catalogs provide us with a parent sample of 1,836 CANDELS galaxies with spectroscopic redshifts at $z_{spec}>0.4$. We remove 35 galaxies that have IRAC colors consistent with AGN using the color selections of \citet{donley12}, leaving a sample of 1,801 galaxies for analysis. The resulting photometric redshifts are shown in Figure~\ref{fig:photz_vs_specz_CANDELS} and are in excellent agreement with the spectroscopic redshift, with $\sigma_\mathrm{NMAD}=0.027$, $\eta=4.8\%$, and $z\text{-}\mathrm{bias}=-0.009$. These results are similar even if we adopt only the same filter set as COSMOS2015 for the CANDELS data, for which we find $\sigma_\mathrm{NMAD}=0.029$, $\eta=4.9\%$, and $z\text{-}\mathrm{bias}=-0.010$. The redshift bias of the CANDELS data is noticeably lower at higher redshifts, with $z\text{-}\mathrm{bias}(z>2)=-0.003$, in contrast to what is found when using the COSMOS2015 catalog ($z\text{-}\mathrm{bias}(z>2)=-0.019$). We attribute these changes to possible differences in adopted zeropoint magnitudes and/or methodology for the photometry between the datasets. Indeed, when we examine overlapping spectroscopic sources between the CANDELS and COSMOS2015 catalogs ($N=741$), we find that the photometry can vary by $10-20$\% in most bands. However, the flux in the shortest wavelength bands ($u$, $IA427$, $B$, etc.) bands are systematically higher ($\sim$10\%) relative to longer wavelength data in the COSMOS2015 catalog when compared to the CANDELS data. This situation is consistent with the slightly lower estimates for \zphot\ (this effect is equivalent to less IGM absorption) when using the COSMOS2015 photometry.

As a separate independent check, we also tested \magphysz\ on other CANDELS fields. We utilize the photometric catalogs available for the GOODS-S \citep{guo13}, UDS \citep{galametz13}, and EGS \citep{stefanon17} fields, together with the spectroscopic redshifts of these fields from the MOSDEF \citep{kriek15}, VANDELS \citep{mclure18a, pentericci18}, and VUDS \citep{tasca17} surveys. We only consider galaxies with `robust' spectroscopic redshifts (reliability $\ge95$\%) which are taken to be objects with \texttt{Z\_QUAL}$\ge5$ in MOSDEF, \texttt{zflg}=3 or 4 in VANDELS, and \texttt{zflag}=3 or 4 in VUDS. Duplicate cases are removed, with a preference towards higher quality cases. For galaxies at $z_{spec}>0.4$, we obtain 370 objects in GOODS-S (51 from MOSFIRE, 185  from VANDELS, 142 from VUDS; 8 duplicates), 264 objects in UDS (30 from MOSFIRE, 235 from VANDELS; 1 duplicate), 433 objects in EGS (451 from MOSFIRE; 18 duplicates). The resulting photometric redshifts are shown in Figure~\ref{fig:photz_vs_specz_CANDELS} and are also in excellent agreement with the spectroscopic redshift, with $\sigma_\mathrm{NMAD}=0.045$, $\eta=5.2\%$, and $z\text{-}\mathrm{bias}=0.006$ ($z\text{-}\mathrm{bias}(z>2)=0.004$).

From these results, we conclude that the \zphot\ values found are subject to change slightly as a result of adopted photometric zeropoints or methodologies and also with the filters that are utilized. It is also important to mention again that intermediate-band data may also impact the accuracy of some sources due to the lack of emission lines in our models. However, the changes in \zbias\ between all samples examined appear relatively small and should not significantly affect the resulting physical properties derived with \magphysz. 

\end{document}